\documentclass{bmcart}

\usepackage[utf8]{inputenc} 

\usepackage{url}
\usepackage{soul}
\usepackage{xcolor}

\usepackage{graphicx}

\startlocaldefs
\endlocaldefs

\usepackage{dcolumn}
\newcolumntype{d}[1]{D{.}{.}{#1}}

\usepackage{placeins}

\makeatletter
\AtBeginDocument{%
  \expandafter\renewcommand\expandafter\subsection\expandafter
    {\expandafter\@fb@secFB\subsection}%
  \newcommand\@fb@secFB{\FloatBarrier
    \gdef\@fb@afterHHook{\@fb@topbarrier \gdef\@fb@afterHHook{}}}%
  \g@addto@macro\@afterheading{\@fb@afterHHook}%
  \gdef\@fb@afterHHook{}%
}
\makeatother

\begin{document}

\begin{frontmatter}

\begin{fmbox}
\dochead{Research}


\title{Novelty and Cultural Evolution in Modern Popular Music}


\author[
   addressref={aff1},                   
   email={kmarie.otoole@gmail.com}   
]{\inits{KO}\fnm{Katherine} \snm{O'Toole}}
\author[
   addressref={aff1},
   email={kmarie.otoole@gmail.com}
]{\inits{E\'AH}\fnm{Emőke-Ágnes} \snm{Horvát}}


\address[id=aff1]{
  \orgname{Northwestern University}, 
  \street{2240 Campus Drive},                     %
  \postcode{60208}                                
  \city{Evanston},                              
  \cny{USA}                                    
}



\end{fmbox}


\begin{abstractbox}

\begin{abstract} 
The ubiquity of digital music consumption has made it possible to extract information about modern music that allows us to perform large scale analysis of stylistic change over time. In order to uncover underlying patterns in cultural evolution, we examine the relationship between the established characteristics of different genres and styles, and the introduction of novel ideas that fuel this ongoing creative evolution. To understand how this dynamic plays out and shapes the cultural ecosystem, we compare musical artifacts to their contemporaries to identify novel artifacts, study the relationship between novelty and commercial success, and connect this to the changes in musical content that we can observe over time. Using Music Information Retrieval (MIR) data and lyrics from Billboard Hot 100 songs between 1974-2013, we calculate a novelty score for each song's aural attributes and lyrics. Comparing both scores to the popularity of the song following its release, we uncover key patterns in the relationship between novelty and audience reception. Additionally, we look at the link between novelty and the likelihood that a song was influential given where its MIR and lyrical features fit within the larger trends we observed.

\end{abstract}




\end{abstractbox}
%

\end{frontmatter}



\section{Introduction}
When NWA dropped their hit song, `Straight Outta Compton', it was one of the hottest new tracks of 1988, but in fact, a key component of the song hearkened all the way back to 1969. By incorporating a sample of the famous `Amen Break' drum solo from The Winstons' song `Amen Brother', the song is an example of the way that musical traits can persist even as they undergo change and reinvention. This juxtaposition highlights the paradox that make culture so fascinating; it provides us with a foundation of established aesthetics and practices to draw from, even as it continues to change and evolve. This dynamic balance between established norms, and the introduction of novelty provides a rich area of inquiry for cultural analysis that looks not only at the impact of novelty on patterns of commercial production and consumption, but also at how the introduction of novel creative artifacts drives cultural evolution \cite{mauch_evolution_2015, weis_investigating_2019, serra_measuring_2012, bomin_evolution_2016, prockup_modeling_2015, klimek_fashion_2019}. With the rise of digital media, information about the consumption and production of cultural artifacts is available to us at unprecedented scale. In addition, the digitization of artifacts allows us to apply computational analyses to better understand the often nebulous concepts of creativity and novelty, and unlock insights into their effects on cultural change. With music in particular, the availability of digital data, and advances in computational methods of audio analysis have made it possible to investigate these questions at scale. The ubiquity of popular music also means that established markers of success, such as Billboard charts, are based on the opinions of a large population. Additionally, while music styles vary across different genres and cultures, music is a `human universal' found in virtually all societies, and displays organizational properties that we can track over time in order to analyze patterns of cultural change \cite{mauch_evolution_2015, serra_measuring_2012}.

Currently it is possible to extract quantitative metrics from large music data sets using Music Information Retrieval (MIR) software. This data, referred to as audio features, or audio descriptors, is information that can be extracted from audio signals, and can be roughly classified as low-level and high-level features. Low-level features directly describe the audio signal data, for example spectral descriptors, while high-level features typically describe more holistic information about the song such as key, energy level, or danceability \cite{magron_leveraging_2020, moffat_evaluation_2015}. Previous work has demonstrated that these high level MIR features provide accurate and robust data for modeling musical preference \cite{magron_leveraging_2020, friberg_using_2014}, along with comparisons of content similarity that inform automatic genre classification \cite{bertin-mahieux_large-scale_2013, lippens_comparison_2004}. This has enabled researchers to contextualize songs within the larger musical ecosystem they exist in, and to identify long term trends in how genres and styles evolve over time  \cite{ mauch_evolution_2015, serra_measuring_2012, weis_investigating_2019, bomin_evolution_2016, prockup_modeling_2015, interiano_musical_2018}. This data is also used by streaming services such as Spotify to develop music discovery tools and generate recommendations for users. In addition to MIR data, word and document embeddings of lyrics have also been shown to be a rich source of data for music content analysis, including genre and mood classification \cite{mayerl_comparing_2020, mayer_combination_2008, hu_lyric_2009, hu_improving_2010, mcvicar_lyric_2021}. As with MIR features, these vector embeddings can be used to evaluate the similarity between lyrics of different songs \cite{mikolov_efficient_2013, le_distributed_2014, whalen_patent_2020}.

In combination with data on commercial success and popularity, MIR and lyric data has enabled researchers to examine how the novelty of songs correlates with their success. For both MIR features and lyrics, the novelty of a song relative to its peers has been found to play a role in determining its cultural success, with the most popular songs demonstrating an optimal level of differentiation that allows them to stand out without being perceived as too dissimilar \cite{askin_what_2017, askin_cultural_2014, berger_are_2018}. Identifying these patterns of optimal differentiation in consumer preferences is important for both the music industry at large, and for development of recommender systems \cite{anderson_algorithmic_2020}. However, previous work has still looked at MIR and lyrical novelty separately, and there is a lack of understanding as to how the relationship between these two dimensions of novelty might affect listeners perception of overall song novelty, and the success of the song. Work in genre and mood classification has shown MIR and lyric data to be complementary, with the inclusion of both sets of features having a positive impact on classification accuracy \cite{  mayerl_comparing_2020, hu_improving_2010, laurier_multimodal_2008, amati_integration_2007}. Since both of these components contribute to the overall perception of the song's mood and genre, we propose to study whether there is also a relationship between a song's MIR novelty and its lyrical novelty, and if that relationship influences its performance on the Billboard Hot 100 chart. Additionally, we consider an alternative definition of success in terms of how likely it was that the song exerted some degree of stylistic influence on the cultural ecosystem. This has been done in previous studies with classical music by tracking the reappearance of specific motifs or harmonic patterns, however the granularity required in this type of analysis makes it difficult to scale \cite{bomin_evolution_2016}. By using MIR and lyric features though, it is possible to perform this type of analysis at scale by using the similarity measures between a song and later releases to determine the likelihood that the song in question was influential \cite{saleh_toward_2016}. In doing so, we can examine whether a song's novelty and initial success affect its likelihood of being influential in the long term, and gain insight into how the introduction of novel attributes fuels ongoing creative evolution in modern popular music. 

In this paper, using MIR and lyric feature data from Billboard Hot 100 songs between 1974-2013, we calculated novelty scores for each song relative to its genre and release year, and compared these to the total number of weeks the song spent on the Hot 100 chart. We found that the novelty scores at which optimal differentiation occurred were quite similar for both MIR and lyrics, and the most successful songs where those that were optimally differentiated for both. When looking at the probability of a song being influential, we also observed optimal differentiation occurring with respect to the novelty scores. Additionally, we found that there was no correlation between the time the song spent on the chart, and its probability of being influential. Rather, we found that for different novelty scores, the amount of time the song spent on the chart affected its likelihood of being influential. By utilizing computational data and methodology to extract high level patterns of change within the musical ecosystem, this research highlights the importance of considering alternate metrics for evaluating success when studying cultural artifacts by providing insight into how novelty affects both short and long term performance of cultural artifacts.

\section{Related Literature}

\subsection{Novelty Metrics}
The word novelty is used to describe ideas or artifacts which are new, original, and in some way dissimilar and different to what came before \cite{uzzi_atypical_2013, li_analyzing_2022}.The production of novelty is important for innovation. Thus, understanding how novelty occurs is a salient question across many domains \cite{liu_hot_2018}, including the sciences \cite{li_analyzing_2022, shin_scientific_2022, shi_weaving_2015}, and creative industries such as music \cite{miles_what_2021}, film \cite{sreenivasan_quantitative_2013}, fashion \cite{klimek_fashion_2019}, and literature \cite{jing_sameness_2019}. Novelty can be evaluated in terms of how similar or dissimilar an artifact is when compared to other artifacts within the larger cultural context, which allows examining its relationship to the cultural space it is embedded in \cite{park_novelty_2020}.

One way this can be achieved is by constructing feature representations that capture information about the key attributes of the artifacts. By representing each artifact based on set of features, this allows us to map individual artifacts to a shared multidimensional feature space and compare them to one another based on their relative positioning. It is then possible to use a distance metric as a way of measuring how similar or dissimilar artifacts are from one another \cite{ mikolov_efficient_2013, le_distributed_2014, whalen_patent_2020, askin_what_2017, askin_cultural_2014,  berger_are_2018, jing_sameness_2019, liu_pandemics_2022}. 

Previous work with music similarity has utilized MIR features to model individual songs as feature vectors due to the ability of MIR features to capture perceptually relevant audio information that has been validated against human perceptions of audio similarity \cite{askin_what_2017, askin_cultural_2014, cheng_exploring_2020}.  

This feature representation approach can also be applied to textual data. Previous research into lyrical novelty used Latent Dirichlet Allocation (LDA) to identify latent topics based on word co-occurence, and represent individual songs based on their topic composition \cite{berger_are_2018}. A similar approach using LDA has also been applied to research into fan fiction, with the Jensen-Shannon Distance between the topic representations of artifact being used to calculate their relative similarity \cite{jing_sameness_2019}. The development of word embedding models means that text can also be represented as a feature vector. These models use textual training data to analyze word usage and map each word to a position in multidimensional feature space based on the context in which they are used, allowing us to compare the similarity of words by compare the relative positions of their vectors to one another \cite{le_distributed_2014}.

For example, the vectors for the words `sad' and `morose' would be closer to one another in the vector space than the vectors for the words `sad' and `happy'. This can also be extended to map longer text inputs such as paragraphs or entire documents to single vectors. This approach has been leveraged in previous work on the analysis of patent novelty, where document embedding models have been used to generate feature vector representations for patents, allowing the cosine similarity between them to be calculated \cite{whalen_patent_2020}. There are also domain specific models that can be used for document embeddings, such as BioBERT, which as has similarly been used to assess the relative novelty of PubMed Articles by generating document feature vectors \cite{liu_pandemics_2022}.

The benefit of this approach to calculating novelty is that mapping artifacts to a shared feature space allows us to contextualize our measurement of novelty within the larger domain context. A challenge with measuring novelty is that what is considered novel is always changing, and to assess whether or not an artifact is novel, we must contextualize it by comparing it to its contemporaries \cite{interiano_musical_2018,zangerle_hit_2019, moore_taste_2013}. By mapping artifacts to a shared feature space and using distance metrics to evaluate similarity, we are able to identify novel artifacts that are informed by this context, without first needing to identify specific markers of novelty.

\subsection{Novelty and Success}

Previous research from psychological studies of culture has suggested that the novelty of cultural artifacts impacts how favorably they will be perceived by audiences \cite{berger_are_2018}. At the individual level, the relationship between subjective novelty and enjoyment has been modeled as an inverse U-curve \cite{berlyne_novelty_1970, chmiel_back_2017}. In this model, objects that are too familiar or too novel will be less successful, with there being a `comfort zone' that describes the desired amount of novelty. When considering large scale consumption, we also see that competition for audience attention means that artifacts needs some degree of novelty to stand out, however audiences are also shown to be averse to very high levels of novelty as well \cite{chai_breakthrough_2019}. In studies of scientific research, a bias against highly novel work has been observed, with very novel work being less likely to be initially recognized and successful, even in cases where high levels of success are achieved in the long term \cite{wang_bias_2017}. Multiple sociological studies exploring this idea across other domains have also found that there appears to be a certain degree of novelty that allows individual artifacts to stand out from their peers, referred to as `optimal differentiation' \cite{ askin_what_2017, askin_cultural_2014}. The idea behind optimal differentiation in the music industry is that although songs must be similar enough to previous work to maintain cohesion in the cultural schema, there must be the introduction of new elements that innovates on the established genre norms and sets them apart, without straying too far out of that comfort zone.

Within the music industry, there are of course many factors that influence the likelihood of a song's success. While it is impossible to control for all of these, previous work has demonstrated that the relationship between an artifact's novelty and their likelihood of success is still significant. Previous research on lyric differentiation found that the degree of differentiation had a significant effect on the ranking of a song on the Billboard digital downloads list even when controlling for amount of radio airplay, artist, and specific lyrical topics \cite{berger_are_2018}. Additionally, research on MIR feature differentiation also found that the effect of differentiation on the amount of time a song spent on the Billboard Hot 100 chart remained significant when controlling for artist popularity in terms of how many times the artist had previously charted, genre preferences, and variations in amount of institutional support that artist received based on whether there were with a major or independent music label \cite{askin_what_2017, askin_cultural_2014}. In this paper, we build on this previous research to examine whether we can observe a relationship between lyric novelty and MIR novelty at the individual song level, as well as whether the relationship between these different types of novelty also correlates with patterns of commercial success.

\subsection{Cultural Evolution}
We can also think of success from the perspective of impacting cultural evolution. We know that over time, musical styles and genres evolve, and their defining characteristics change. As novel artifacts are introduced into the wider cultural ecosystem, they bring new ideas and creative perspectives, which may or may not be incorporated into the existing stylistic norms \cite{park_novelty_2020}. Studies have shown that we can quantitatively track this evolution over time by analyzing changes to the presence and frequency of musical features over time \cite{mauch_evolution_2015, weis_investigating_2019, serra_measuring_2012, bomin_evolution_2016}. We can therefore see whether the features of a given artifact become more or less prominent in the style as a whole over time. If later artifacts are very similar to the artifact in question, this tells us that many of the artifact's features have been incorporated into the stylistic norms, and therefore the artifact is more likely to have been stylistically influential. Although it is not possible to prove a causal relationship, measuring the degree of similarity between a cultural artifact and other artifacts produced at a later time is a standard approach for inferring potential influence \cite{saleh_toward_2016}. Although novelty plays an important role in fueling stylistic evolution, we are lacking empirical evidence about the correlation between the degree of novelty in an artifact, and how likely it is that the artifact will be influential.

\section{Research Questions}
Based on the above gaps in the literature, we aimed to answer the following research questions:
\begin{itemize}
    \item \textbf{RQ 1:} What is the relationship between a song's MIR novelty and its lyric novelty?
\item \textbf{RQ 2:} Does the relationship between a song's MIR novelty and lyric novelty impact its likelihood of success?
\item \textbf{RQ 3:} Does the MIR novelty and/or lyric novelty of a song impact the probability of the song being influential?
\item \textbf{RQ 4:} How does the amount of change over time to the average MIR features compare to the amount of change over time to the average lyric features? 

\end{itemize}

\section{Data}

Our data comprises songs from the Billboard Hot 100 chart, which tracks the 100 most popular songs in the United States for each week  based on Nielsen radio play scores, physical and digital music sales, along with streaming figures. It therefore serves to track the commercial success of individual songs. For the purposes of our analysis, we measured the success of each song at the time of their initial release based on how many weeks it had been included on the Hot 100 Chart. More time spent on the chart was therefore indicative of higher degrees of success. The Billboard Hot 100 chart is an industry standard for measuring song popularity, and has been used in numerous studies on popular music due to their reliable insight into the most popular American music at a given time \cite{askin_cultural_2014}. This data set allowed us to limit our analysis to only popular music that was most representative of the prevailing cultural space at each point in time. The data set included song genre from Discogs.com, the total number of weeks each song spent on the chart, and MIR feature data. We acquired text of the lyrics for each song from a variety of online sources with our custom scraping tools. The subset of the data used in our analysis consisted of 14,248 songs that were on the Billboard Hot 100 chart between 1974-2013, encompassing 3,973 unique artists and bands across 643 different record labels and 17 genres.

\section{Methodology}
\begin{figure}[h]
\centering
\includegraphics[width=12cm]{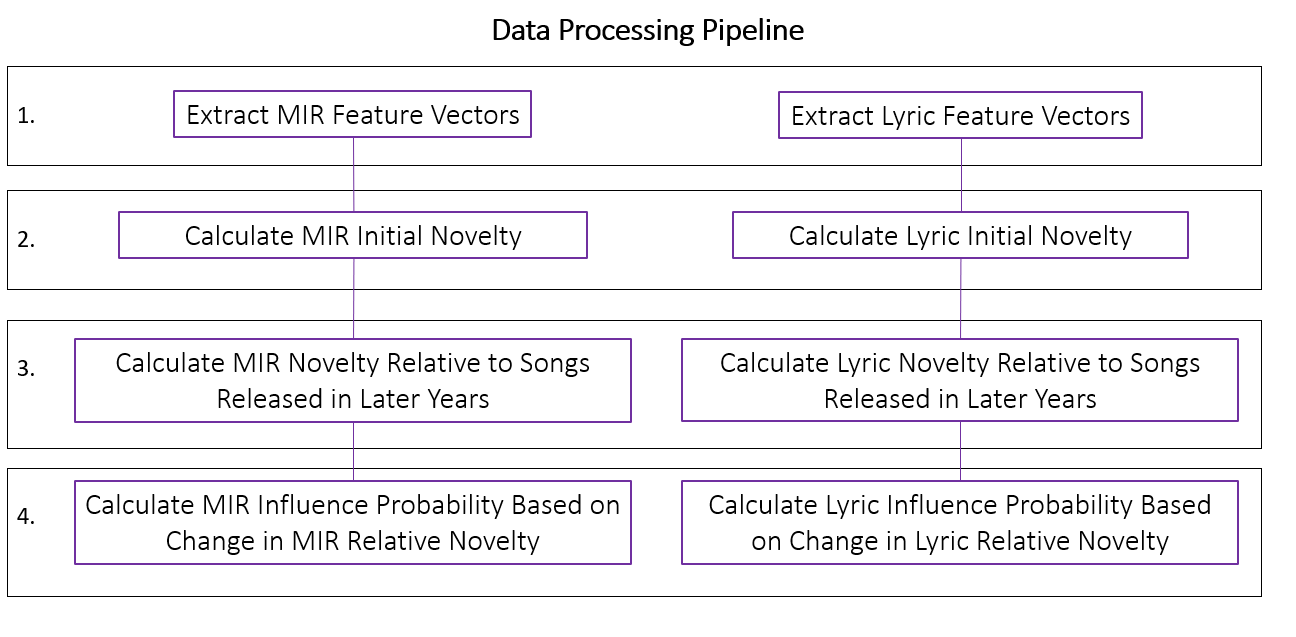}
\caption{Each of the steps in our data processing pipeline used to generate the MIR and lyric song novelty metrics used in our analyses.}
\label{fig:data_pipeline}
\end{figure}

In Figure \ref{fig:data_pipeline} we have included an illustration of the data processing pipeline used to generate the metrics used for our analyses. 

\begin{enumerate}
  \item Extract MIR and Lyric Features: MIR and lyric features represent dimensions that define an MIR feature space and a lyric feature space, respectively. The feature values we derived for each song provide us with a vector that maps that song to these multi-dimensional feature spaces. This allows us to compare the aural and lyrical similarity of sets of songs based on the relative positions of their MIR feature vectors, and the relative positions of their lyric feature vectors. 

\item Calculate Song Novelty: Since the novelty of a song is assessed relative to the other songs released in the same time period, we can group the songs based on the year they were released. We also choose to only generate within-genre novelty comparisons, as the stylistic variation between genres means that cross-genre comparisons would not give us a good measure of novelty. We can then group the songs by genre and release year and calculate an MIR novelty score and a lyric novelty score for each individual song, based on the average distance between the song's vector and the other song vectors that were \textit{released in the same year and genre.} For example, when calculating the novelty of a rock song released in 1985, we would compare it to the other rock songs that were also released in 1985. The farther away an individual song's vector is from the average position of the song vectors in that subset, the more novel that song is.

\item Calculate Song Relative Novelty: Once we have computed the initial novelty score of a song, relative to the year and genre it was released in, we can calculate how novel it is compared to songs that are in the same genre, but were released in a later year. This gives us a score which shows the relative novelty of the song when compared to a given year. For example, if a Rock song was released in 1982, we could calculate its relative novelty with respect to 1985 by finding the average distance between its feature vectors, and the feature vectors of all the Rock songs that were released in 1985.

\item Calculate Influence Probability: We can calculate the change in relative novelty by subtracting the song's relative novelty score from its initial novelty score. If there is an increase in relative novelty, it is not likely that the song was influential. If there is a decrease in relative novelty, it is more likely the song was influential. 

\end{enumerate}

\subsection{Feature Extraction}
The MIR features used in our analysis consisted of quantitative data for 13 high level MIR features which were derived from The Echo Nest using their Music Information Retrieval (MIR) system. A description of the MIR features can be found in Table \ref{table:Echo_Nest} in the Appendix. 

The lyrics features were generated using a document embedding system.  We cleaned, preprocessed, and tokenized the lyrics using the Gensim simple preprocessing utility \cite{radim_rehurek_gensim_2010}. We then trained a Doc2Vec model which had a vector of 100 dimensions and iterated over the training corpus 40 times \cite{le_distributed_2014}. The minimum word count was set to 2 in order to discard words with a single occurrence. This model was then used to generate a 100 length feature vector for the lyrics of each individual song. Unlike the MIR features, the lyric features do not map to concrete concepts, however all together they define a feature space where we can compare how relatively similar the contents of two documents are by looking at how close their vectors are.

\subsection{Novelty Scores}
To generate our novelty scores, we calculated the distance between each individual song vector, and the other song vectors within the same genre and release year. We opted to use a distance metric when quantifying the change in average genre positioning, and the individual song novelty scores, as opposed to cosine similarity which was used in previous studies. This allows us to track how much the average position of a genre's feature vectors is changing over time for both MIR features and lyric features, in addition to measuring the individual song novelty scores. 

For the lyric vector distances, we used Euclidean distance, as there was no significant covariance between any of the individual features that comprised the feature vectors. For each song, the lyric feature vector can be written as: 

\[{\displaystyle {\vec {l}}=(l_{1},l_{2},l_{3},\dots ,l_{100})} \]

For each genre-year group of songs, we can then take the component-wise average across all the individual song vectors to calculate the average feature vector for that genre-year group, which can be written as: 
\[{\displaystyle {\vec {\mu}}=(\mu_{1},\mu_{2},\mu_{3},\dots ,\mu_{100})}\]

We then calculate the Euclidean distance between this average vector, and the individual song vectors for each song within the genre-year group:

\[{\displaystyle d({\vec {l}},{\vec {\mu}})={\sqrt {( l_{1}-{\mu_{1} )^{2} + ( l_{2}-{\mu_{2} )^{2} + \dots + ( l_{100}-{\mu_{100} )^{2}  }}}}} .}\]

For the MIR vector distances, we followed a similar approach.  With MIR data however, some feature values are used as input for determining the values of other features, which means that there are covariances between the features. For example when calculating the danceability of a song, tempo and valence values are included. As a result, we cannot use the Euclidean distance, and instead need to use the Mahalanobis distance. The Mahalanobis distance is similar to Euclidean distance, but is calculated using the covariance matrix of all of the feature vectors, so that it can account for any dependencies between the features, and scale the distance accordingly. 

For each song, the MIR feature vector can be written as: 

\[{\displaystyle {\vec {m}}=(m_{1},m_{2},m_{3},\dots ,m_{13})}\]

We again take the average of all the song vectors within the genre-year group in order to yield the average MIR feature vector:

\[{\displaystyle {\vec {\mu}}=(\mu_{1},\mu_{2},\mu_{3},\dots ,\mu_{13})}\]

We then calculate the covariance matrix, $C$ for the MIR feature vectors, which gives us the covariance between each pair of MIR features included in our feature vectors. In order to be consistent when calculating the within-genre distances for different years, a covariance matrix was generated for each genre using data from all years, and used in the distance calculations, rather than generating a covariance matrix for each genre-year subset. Because covariances between MIR features mainly vary across genres, but are fairly consistent within genre over time, this allowed us to make sure that the normalization applied by the Mahalanobis distance calculation was consistent across all the subsets of a genre, allowing us to compare different time periods. 

The Mahalanobis distance $D_{M}$ between an individual song song ${\vec {m}}$ and the rest of the song vectors in the same genre-year group can then be calculated as follows:

\[{\displaystyle D_{M}({\vec {m}})={\sqrt {({\vec {m}}-{\vec {\mu }})^{T}C^{-1}({\vec {m}}-{\vec {\mu }})}}.}\]

Calculating the individual song distances yields a distribution of MIR vector distances and a distribution of lyric vector distances for each of the genre-year subsets. In Table \ref{table:avg_nov} we have included the total number of songs for each year, as well as the average MIR and lyric vector distances across all genres for each year.

\begin{table}[ht]
\centering
\caption{Number of songs per year and average MIR and lyric vector distances across all genres}
\label{table:avg_nov}
 \begin{tabular}{c | c | c | c } 
\textbf{Year} & \textbf{\# of Songs} & \textbf{Avg. MIR Vector Distance} & \textbf{Avg. Lyric Vector Distance} \\ \hline
1974 	& 464 	&3.45 	&13.45 \\ \hline
1975 	&495 	&3.39 	&13.11 \\ \hline
1976 	&468 	&3.47 	&13.51 \\ \hline
1977 	&412 	&3.42 	&13.27 \\ \hline
1978 	&422 	&3.33	&13.78 \\ \hline
1979 	&435 	&3.30  &14.18 \\ \hline
1980 	&428 	&3.36	&13.52 \\ \hline
1981 	&376 	&3.25	&13.40 \\ \hline
1982 	&397 	&3.32  &13.62 \\ \hline
1983 	&427 	&3.23  &13.83 \\ \hline
1984 	&414 	&3.23 	&14.24 \\ \hline
1985 	&391 	&3.09 	&14.20 \\ \hline
1986 	&384 	&3.11 	&13.99 \\ \hline
1987 	&381 	&3.06 	&14.04 \\ \hline
1988 	&371 	&3.03 	&13.86 \\ \hline
1989 	&382 	&3.08 	&14.09 \\ \hline
1990 	&353 	&3.03 	&14.06 \\ \hline
1991 	&353 	&3.15 	&13.95 \\ \hline
1992 	&344 	&3.21 	&14.12 \\ \hline
1993 	&331 	&3.23 	&14.64 \\ \hline
1994 	&318 	&3.33 	&14.36 \\ \hline
1995 	&339 	&3.18 	&14.38 \\ \hline
1996 	&304 	&3.22 	&14.06 \\ \hline
1997 	&329 	&3.27	&14.10 \\ \hline
1998 	&330 	&3.36 	&14.23 \\ \hline
1999 	&313 	&3.38 	&14.40 \\ \hline
2000 	&313 	&3.28 	&14.57 \\ \hline
2001 	&297 	&3.34 	&14.86 \\ \hline
2002 	&291 	&3.29 	&14.68 \\ \hline
2003 	&302 	&3.41 	&14.60 \\ \hline
2004 	&303 	&3.36 	&14.68 \\ \hline
2005 	&341 	&3.41 	&14.28 \\ \hline
2006 	&361 	&3.31 	&14.52 \\ \hline
2007 	&337 	&3.38 	&14.88 \\ \hline
2008 	&387 	&3.31 	&14.76 \\ \hline
2009 	&429 	&3.30 	&14.71 \\ \hline
2010 	&474 	&3.34 	&14.71 \\ \hline
2011 	&487 	&3.27 	&14.65 \\ \hline
2012 	&369 	&3.17 	&14.75 \\ \hline
2013 	&174 	&3.08 	&14.38 \\ \hline
\end{tabular}
\end{table}

In order to compare song novelty between songs from different years and genres, we then normalize each song's vector distance relative to the mean distance and standard deviation of the distances for all the songs within the same genre-year subset. This allows us account for variations in the distributions of distances within each genre-year subset. We do this by calculating the z-score for each individual vector distance. The z-score tells us the relative positioning of a vector distance within a distribution by subtracting the mean distance of the distribution, and then dividing the difference by the standard deviation of the distribution. In doing so, the z-score tells us how many standard deviations from the mean that particular value is, which indicates how novel the vector distance is, and allows us to compare it to novelty scores drawn from different distributions. 

\subsection{Relative Novelty Scores}
In order to evaluate the likelihood that a song was influential, we have to compare it to the cultural ecosystem at later points in time. To compare a song's similarity to songs released in later years, we can use the same approach that we took for calculating the novelty scores, but instead of comparing the song to songs in the same genre-year subset, we compare it to songs in the same genre, but released in a later year. 

Not all of the genres in our dataset had songs included on the Hot100 chart for every year within the time period we looked at, meaning there were a large number of relative novelty scores that could not be calculated. Because of this,  we limited our analysis to the 6 genres with the greatest number of songs; Rock, R\&B, Rap, Country, Pop, and Electronica. For each song within these genres, we compared it to the genre-year subsets of the subsequent ten years following the song's release. For example, if a Rock song was released in 1982, we would compare its feature vector to the average feature vector of Rock songs that were released in 1983, 1984, 1985 and so on. Using the same process as we used for calculating the initial novelty scores, we calculated the Euclidean distance for the lyric vector distances, and the Mahalanobis distance for the MIR vector distances. Again, in order to account for variations in the mean distance and total range of distances within different genre-year distributions, we calculated the z-score for each of the 10 relative vector distances that had been calculated for each song. For each relative vector distance, this was done using the mean and standard deviation of the distances in the genre-year subset that had been used for that specific relative comparison. This yielded  relative novelty scores that we could then compare to the relative novelty scores for other years, and to the initial novelty score.

\subsection{Influence Probability}

\begin{figure}[h]
\centering
\includegraphics[width=12cm]{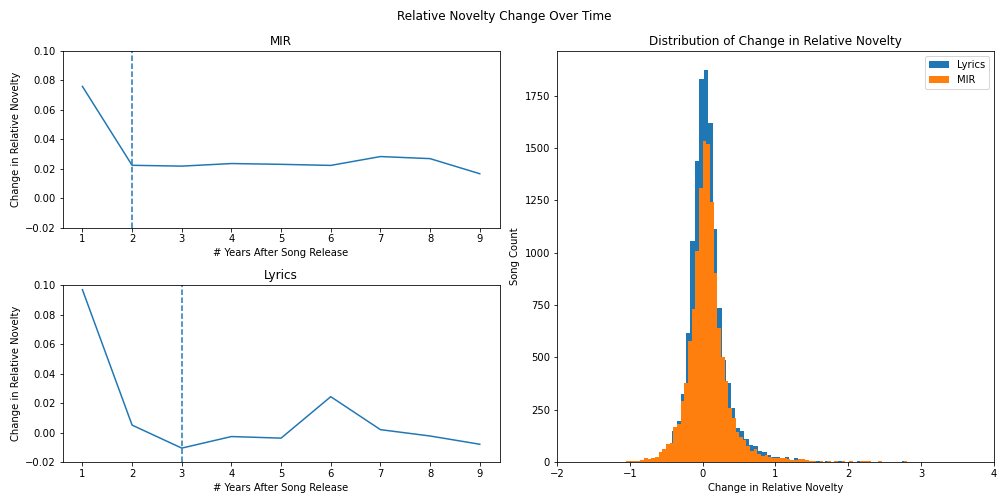}
\caption{The two plots in the left column show the magnitude of year over year change in the average relative novelty score as songs are compared to later release years. For both MIR relative novelty and lyric relative novelty, we see that the rate of change is steep for the first two years after release, then stabilizes. The vertical lines indicate the inflection points where this occurs, which for MIR relative novelty is after two years, and for lyric relative novelty is after 3 years. The plot in the right column shows the distributions of relative novelty change for MIR novelty, and lyric novelty, which are calculated by taking the average change in relative novelty which occurred in the first three years after the song's release.}
\label{fig:rel_nov_change}
\end{figure}

We can determine whether or not it was probable that a song was influential or not based on whether its relative novelty score had increased or decreased in relation to its initial novelty score. Since we evaluate the relative likelihood that a song was influential by calculating the change in both its MIR and lyric relative novelty, we first determined whether the rate at which relative novelty changed was consistent over time. Taking the average change in relative song novelty in the years following its initial release, we found that the rate of change for MIR relative novelty plateaued after 2 years, and the rate of change for lyric relative novelty plateaued after 3 years (see Figure~\ref{fig:rel_nov_change} left plots). Because of this, we decided to only consider the relative novelty change that occurred in the 3 years following a song's release by taking the average of the relative novelty scores for those first three years, and subtracting the song's initial novelty score. Songs released in 2011 or later were excluded since we did not have data for the full three years following those release years. For example, for Prince's, `When Doves Cry', the average change in MIR relative novelty was an increase of 0.15, and for lyric relative novelty it was an increase of 0.12. Because the relative novelty increased, this tells us that between 1985 and 1987, the average MIR and lyric features of the Rock genre became more dissimilar to the features seen in `When Doves Cry'.

\subsection{Average Feature Change Over Time}
In addition to the initial novelty and relative novelty of individual songs, we also consider the magnitude of both MIR and lyric stylistic change over time. Within our data set, this can be understood as changes to the average position of the genre's feature vectors in feature space over time. Since a snapshot of a genre's position at a given time is represented by a distribution of song feature vectors, to consider the novelty score within a broader context of the genre's movement in feature space, we examined the amount of feature variance seen within an individual genre over time for both MIR and lyric features. To perform this analysis, we assigned each song to a decade based on their release years, grouping them into ten-year intervals of 1974-1983, 1984-1993 and so on. Since the MIR and lyric feature vectors for each song provide us with an attribute-based representation of our data, we can compare how distinct the feature distributions of each class are from one another by training a decision tree classifier to predict the temporal class of a given song based on its feature values. Using the training data, a decision tree learns how to partition the feature space to best predict the temporal class of a song. The more distinct the area of feature space that each class inhabits, and the less overlap each has with other classes, the more accurate the decision tree. As a result, a higher accuracy tells us that there is less similarity between the feature distributions of different decades.

For our classifier, we used the random forest classifier model from the scikit-learn library \cite{pedregosa_scikit-learn_2011}. A random forest works by fitting multiple decision trees to the data, and averaging results to improve accuracy and avoid overfitting. For our model, we used 500 trees with no depth limit. For individual genres, we then compared the accuracy of the classifier in predicting the temporal class of individual songs when trained using the MIR features versus when trained using the lyric features. For each set of features, a cross-validation was run using a repeated K-fold with 5 splits and 5 repeats, allowing us to generate the distribution of accuracy scores across different train-test splits of the data.

\section{Results}

\subsection{Relationship Between MIR Novelty and Lyric Novelty}
\begin{figure}[h]
\centering
\includegraphics[width=10cm]{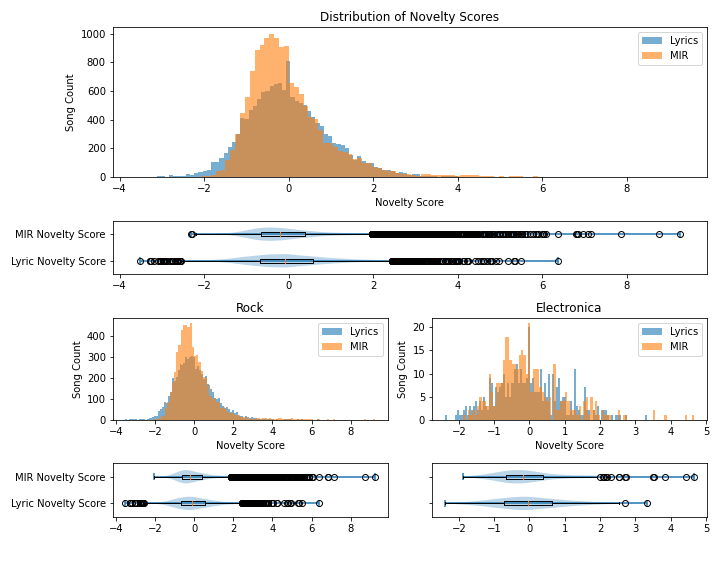}
\caption{Distribution of MIR novelty scores and lyric novelty scores. Top plot displays the distributions for all songs within the dataset. The bottom two plots display the distributions for all songs within the Rock genre and Electronica genre, respectively.}
\label{fig:novelty_distribution}
\end{figure}

In comparing the MIR and lyric novelty score distributions across all years and genres, we found that the MIR novelty distribution had a greater positive kurtosis and a greater positive skew than the lyric novelty distribution (see Fig. \ref{fig:novelty_distribution} top plot and Table~\ref{table:novelty_distributions}). This tells us that there is a greater range in the above average novelty scores occurring within the MIR distribution. Although we can observe that the median value for the MIR novelty distribution, -0.21, is slightly lower than the median value for the lyric novelty distribution, -0.09, a one-way ANOVA test confirms no significant difference between the MIR novelty distribution and the lyrics novelty distribution (F=6.11e-30, p=1.0). These trends held true when analyzing the novelty distributions within individual genres (Fig. \ref{fig:novelty_distribution} bottom plots and Table~\ref{table:novelty_distributions}). 

\begin{table}[h]
\centering
\caption{Novelty Score Distributions}
\label{table:novelty_distributions}
 \begin{tabular}{| p{0.3\linewidth} | c | c|} 
   \hline
\textbf{Modality} & \textbf{Skew}& \textbf{Kurtosis}
 \\ \hline
MIR Novelty - All Genres & 1.88   &   6.53 
 \\ \hline
Lyric  Novelty - All Genres 	&	0.60	&   1.13
 \\  \hline
 
 MIR Novelty - Rock & 2.21	&    8.88
 \\ \hline
Lyric  Novelty - Rock 	&	0.69	&   1.48
 \\  \hline
 
 MIR Novelty - Electronica & 1.32	&     2.82
 \\ \hline
Lyric  Novelty - Electronica 	&	0.33	&   -0.05
 \\  \hline

\end{tabular}
\end{table}

When examining the relationship between MIR novelty and lyric novelty of individual songs, we did not find a significant correlation between the two (Pearson correlation test r=-0.01, p=0.10). They appear to be independent of one another, with no consistent patterns found in the relationship between the MIR novelty score and the lyric novelty score of a given song. As a result, a set of songs with the same MIR novelty score might have a wide variation in their lyric novelty scores, and vice versa. 

\subsection{Initial Success}
We incorporated song commercial success data to determine whether these variations in the novelty distributions of the two modalities were indicative of differences in how they impacted the likelihood of a song becoming popular. Using total number of weeks on chart as the metric for song success, we looked at the success of individual songs in relation to their MIR novelty score and the lyric novelty score. For example, Prince's 1984 song, `When Doves Cry', spent 21 weeks on the chart, and had an MIR novelty score of -0.63, putting it at the 26th percentile, and a lyric novelty score of 0.48, putting it at the 68th percentile. This tells us that the MIR features of the song were less novel when compared to other rock songs in 1984, but that the lyric features were more novel.

\begin{figure}[h]
\centering

\includegraphics[width=12cm]{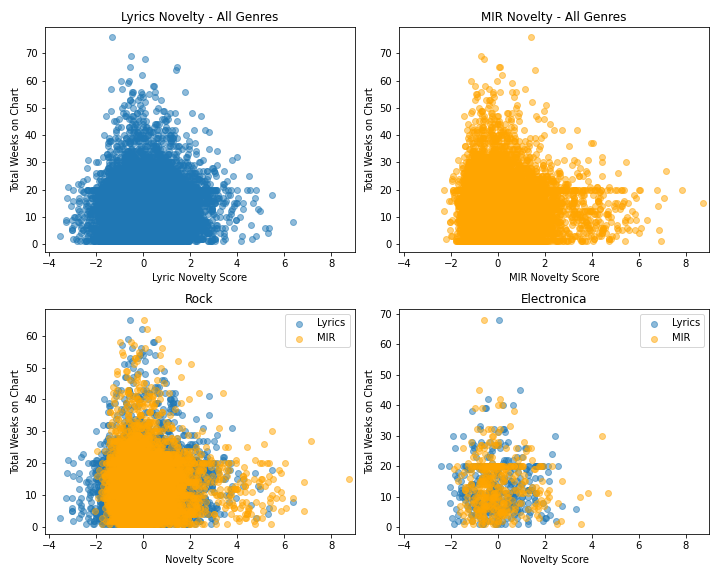}
\caption{Joint distributions of song novelty scores and total number of weeks the song spent on the chart. Top row shows the joint distributions for all genres. Bottom row shows the overlay of MIR-total weeks joint distribution and lyrics-total weeks joint distribution for Rock, and Electronica. In each distribution we can see that the highest number of total weeks on chart occur within a certain range of novelty scores.}
\label{fig:differentiation}
\end{figure}

We found that similar to previous findings, the most popular songs had a degree of optimal differentiation both for MIR novelty, and for lyric novelty \cite{askin_cultural_2014, berger_are_2018} (see Fig. \ref{fig:differentiation} top row). We looked at the relationship between novelty and success for each modality separately, and using the Hotelling T2 test, found no statistically significant difference between the joint distribution of total weeks on chart with respect to MIR novelty, and the joint distribution of total weeks on chart with respect to lyrics novelty (F=3.07e-30, p=1.0). This was also found to be the case at the genre level as well (Figure~\ref{fig:differentiation} bottom row). Specifically, Hotelling T2 test found no significant difference between the MIR joint distribution and the lyrics joint distribution for either Rock (F=7.03e-30, p=1.0) or Electronica (F=5.44e-31, p=1.0).

Given that the songs with the most success fell into a rather narrow range of novelty values, we performed a Kernel Density Estimation analysis to estimate both the MIR novelty score and lyric novelty score which had the highest probability of being in the top 85th, 90th, and 95th percentile of total weeks on chart.For both the MIR-total weeks joint distribution and lyrics-total weeks joint distribution, the Python library scikit-learn was used to generate a Kernel Density Estimation using a Gaussian mixture model and a bandwidth of 0.3. \cite{pedregosa_scikit-learn_2011} The KDE was used to generate probability scores for hypothetical pairings of novelty scores and total weeks on chart, which indicated the likelihood that a song with the given novelty score would be on the chart for the given number of weeks. This was done for 250,000 individual generated data points that were equally distributed across 500 unique values in the range of -1 to 1, which represented novelty values, and across 500 unique values in the range of 20 to 76, which represented the top 85th percentile of total weeks on chart. For each novelty value, we took the summation of the generated probability scores to calculate the relative probability that a song with that amount of novelty would reach anywhere within the top 85th percentile of total weeks on chart. This process was then repeated for the top 90th percentile of total weeks on chart, and the top 95th percentile of total weeks on chart. 

\begin{figure}[h]
\centering
\includegraphics[width=12cm]{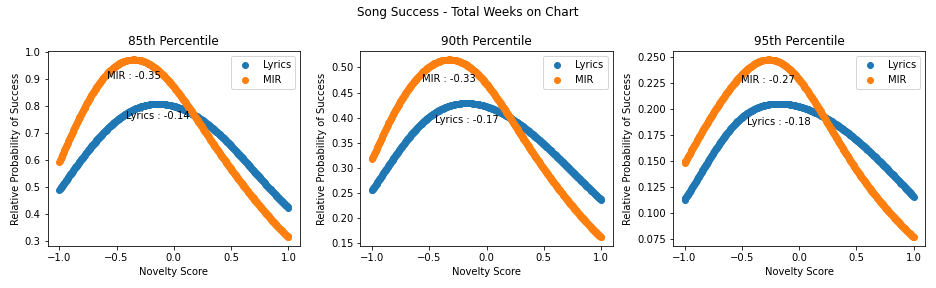}
\caption{Relative probability of success in reaching the 85th, 90th, and 95th percentile of total weeks on chart across a range of MIR novelty scores, and across a range of lyric novelty scores. Since the relative probability values are a summation of probability estimates of a generated data set, they are dependent on the total number of data points generated for our sample, and should not be treated as an absolute probability value.}
\label{fig:prob_distribution_all}
\end{figure}

For each of these, we can see in  Fig. \ref{fig:prob_distribution_all} that for both MIR novelty and lyric novelty, the probability of success increases as the novelty score increases, until a certain point at which it peaks and then because to decrease again. The novelty score for this peak value that we have estimated in our analysis indicates the degree of optimal differentiation that is most likely to help them succeed. Below this, the song is likely to be too similar to stand out from other songs, while above this, it starts to diverge too much from what the audience expects. While the novelty scores of a song cannot be used to predict exactly how successful it will be, songs with novelty scores close to our estimates will have a greater chance of achieving high levels of success than songs with novelty scores that are higher or lower. 

We found that for both lyrics and MIR, the novelty scores which had the highest probabilities of success for each of these performance tiers was just slightly lower than the mean novelty scores of the population. We also found that across the three total week ranges we tested, the  MIR novelty score was consistently slightly lower than the lyric novelty score (see Fig. \ref{fig:prob_distribution_all} and Table~\ref{table:novelty_distributions_success}). Additionally, as the analysis narrows from the top 85th percentile to only the top 95th percentile, we also see that that the MIR novelty score increases, while the lyric novelty score decreases, causing the difference between them to grow smaller. 

\begin{table}[ht]
\centering
\caption{Novelty Score with Highest Success Probability}
\label{table:novelty_distributions_success}
 \begin{tabular}{|p{0.3\linewidth} | c | c| c|} 
  \hline
\textbf{Total Weeks Range} & \textbf{85th Percentile}& \textbf{90th Percentile}& \textbf{95th Percentile}
 \\ \hline
MIR Novelty Score &	-0.35&  -0.33 & -0.27 
 \\ \hline
Lyric Novelty Score & -0.14	& -0.17	& -0.18   
 \\  \hline
\end{tabular}
\end{table}

Given that we did not find any consistent patterns in the relationship between the MIR novelty and lyric novelty of individual songs, we wanted to explore whether different combinations of MIR novelty scores and lyric novelty scores would impact a song's probability of success. For this, we generated a Kernel Density Estimation for the joint distribution which included both lyric novelty and MIR novelty, along with total weeks on chart. This was then used to calculate the probability scores for 1,000,000 equally distributed generated data points having lyric novelty scores between -1 to 1, MIR novelty scores between -1 to 1, total weeks on chart in the 90th percentile, between 22 to 76. For each unique pair of MIR and lyric novelty scores, we took the summation of the generated probability scores to calculate the relative probability of a song with those scores being in the top 90th percentile of total weeks on chart.

\begin{figure}[]
\centering
\includegraphics[width=10cm]{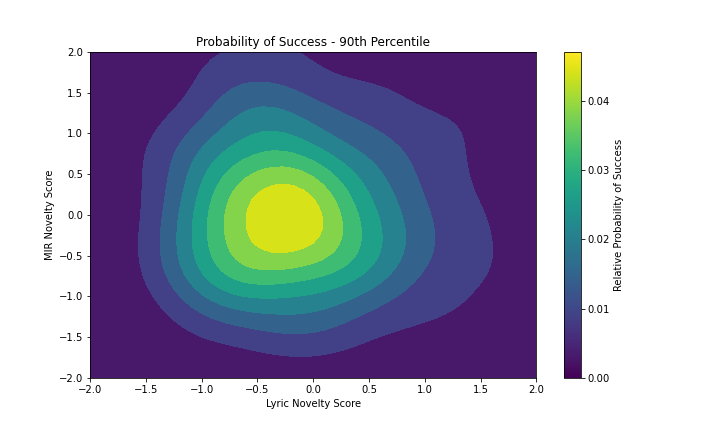}
\caption{Relative probability of success in reaching the 90th percentile of total weeks on chart for different combinations of MIR and lyric novelty scores. We see that the probability decreases as either score moves away from the value that provides the optimal degree of differentiation within its modality. Since the relative probability values are a summation of probability estimates of a generated data set, they are dependent on the total number of data points generated for our sample, and should not be treated as an absolute probability value.}
\label{fig:prob_distribution_both}
\end{figure}

We see in Fig. \ref{fig:prob_distribution_both} that the highest relative probability of success occurs when a song is close to the optimal novelty values for both lyrics and MIR. As we move outward from this area where both novelty scores are close to optimal, the radial pattern of the gradient indicates that variance in the probability of success is equally affected by both variance in the MIR novelty score, and variance in the lyric novelty score. Additionally, given that the gradient is roughly equal for points that have the same distance from this optimal center point, this tells us that the proportional relationship of MIR novelty to lyric novelty is not an explanatory variable, but rather it is the combined total distance from the optimal center that impacts a songs probability of success.

Whether the point lies above or below the optimal value for either novelty score does not change the relationship, which tells us that having a higher than optimal novelty score for one modality can't be 'balanced out' by a lower than optimal novelty score for the other modality. If that were the case, and it were only about reaching an optimal value for the average of the the two novelty scores, then we would expect to see, for instance, a song with an MIR novelty score of -0.22, 0.1 higher than optimal, and a lyric novelty score of -0.27, 0.1 less than optimal, to have an equal probability of success as a song with the optimal values of an MIR novelty score of -0.32 and a lyric novelty score of -0.17. In the contour map we see that the MRI novelty score that optimizes success probability does not change for different values of lyric novelty, and vice versa, the lyric novelty score that optimizes success does not change for different values of MIR novelty. However, we also observe that the combined distance of both novelty scores from their respective optimal values will impact a song's probability of success, indicating that even though a song's lyric novelty and MIR novelty are independent of one another, their deviations from the optimal values will have an additive effect on the overall perception of song novelty as it relates to optimal differentiation.

\subsection{Influence Probability}
It is worth noting that when looking at the inflection points in  Figure~\ref{fig:rel_nov_change} that indicate a plateau in the rate of change of relative novelty, the average year over year change in relative novelty for MIR stays positive, meaning that on average, the MIR features of a genre will tend to become less similar to those in previous years as the gap in time increases. The average year over year change in relative novelty for lyrics, however, is for the most part negative, suggesting that song lyrics within a genre tend to be more similar to those of songs released in previous years. When comparing the distribution relative novelty change for both modalities, we see that both follow a normal distribution, with a one-way ANOVA test showing no statistically significant difference between them (F=0.05, p=0.83). However, the distribution of lyric relative novelty change shows a larger positive kurtosis and skew than that of the MIR relative novelty change distribution, which is the opposite of the trend we observed between MIR and lyrics in the initial novelty score distributions  (see Fig. \ref{fig:rel_nov_change} right plot and Table~\ref{table:Influence_dist}). Additionally, while the difference is not statistically significant, we observe greater variance between the two relative novelty change distributions than we do between the initial MIR and lyric novelty score distributions.

\begin{table}[h]
\centering
\caption{Relative Novelty Change Distributions}
\label{table:Influence_dist}
 \begin{tabular}{| p{0.3\linewidth} | c |c|} 
  \hline
\textbf{Modality} & \textbf{Skew}& \textbf{Kurtosis}
 \\ \hline
MIR Relative Novelty & 6.77   &   123.95
 \\ \hline
Lyric Relative Novelty 	& 15.18	&  450.16
 \\  \hline
 
\end{tabular}
\end{table}

\begin{figure}[h]
\centering
\includegraphics[width=12cm]{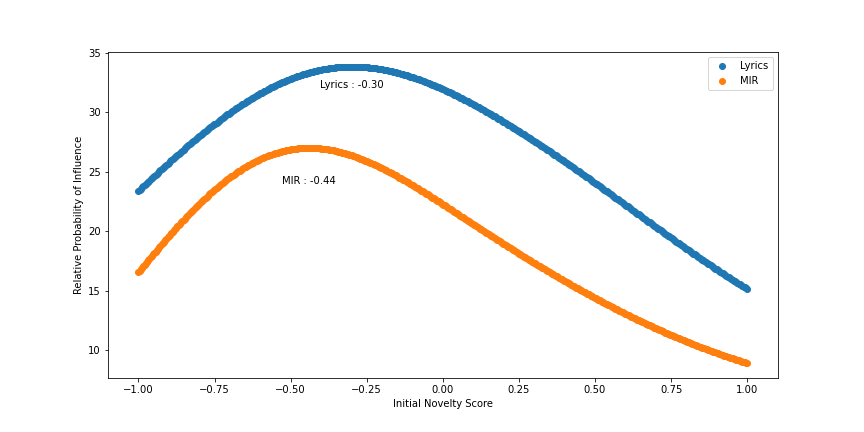}
\caption{Relative probability of a song being in the bottom 10th percentile for relative novelty change based on initial novelty scores. We see that the probability decreases as either score moves away from the value that provides the optimal degree of differentiation within its modality. Since the relative probability values are a summation of probability estimates of a generated data set, they are dependent on the total number of data points generated for our sample, and should not be treated as an absolute probability value.}
\label{fig:novelty_influence_optimal}
\end{figure}

Since a greater decrease in a song's relative novelty indicates a higher likelihood that the song was influential, scores in the bottom 10th percentile represent high performers for this metric. To determine whether a song's initial novelty scores had any correlation with how its relative novelty changed over time, we generated a Kernel Density Estimation for the joint distribution between MIR novelty scores and MIR relative novelty change, as well as for the joint distribution between lyric novelty scores and lyric relative novelty change. Using the same procedure as for the initial success KDE, we found that for both lyrics and MIR, the novelty scores that correlated with the highest probabilities of seeing a large decrease in relative novelty were below the average novelty score of the population, and slightly lower than than the optimal differentiation novelty values we estimated for initial success (see Fig. \ref{fig:novelty_influence_optimal}). Again, we see that the optimal MIR novelty score is slightly lower than the optimal lyric novelty score.

Because optimal differentiation is about the relationship between cultural artifacts and their contemporaries, the fact that we see a degree of optimal differentiation in the relationship between artifacts released at different points in time is a new finding. While one possible explanation is that the initial popularity of songs that are optimally differentiated makes them more likely to be influential, we did not find any correlation between total number of weeks on chart, and either lyric or MIR relative novelty change. For example, while both the 1998 Janet Jackson song `Together Again' and the 2005 Kelly Clarkson song `Since U Been Gone' were on the chart for a total of 46 weeks, the relative MIR novelty for `Together Again' decreased by -0.21, while the relative MIR novelty for `Since U Been Gone' increased by 0.22. These results suggest that the amount of time a song has spent on the chart cannot be used to predict its likelihood of being influential.

To investigate this, we ran an analysis to examine the relationship between total weeks on chart and influence probability when controlling for limited ranges of novelty scores. Running two analyses, one for MIR novelty and MIR relative novelty change, and one for lyric novelty and lyric relative novelty change, we considered songs that fell within three different ranges of novelty scores; the bottom 10th percentile, the top 10th percentile, and the 10th percentile centered around the novelty score with the highest probability of maximizing total weeks on chart. Within each range, the songs were then grouped by the number of weeks they had spent on the chart. An aggregated influence probability was calculated for each week in the range of 0 to 35 by calculating the percentage of songs in that grouping whose relative novelty had decreased. Additional details for this process can be found in the Appendix.

\begin{figure}[ht]
\centering
\includegraphics[width=12cm]{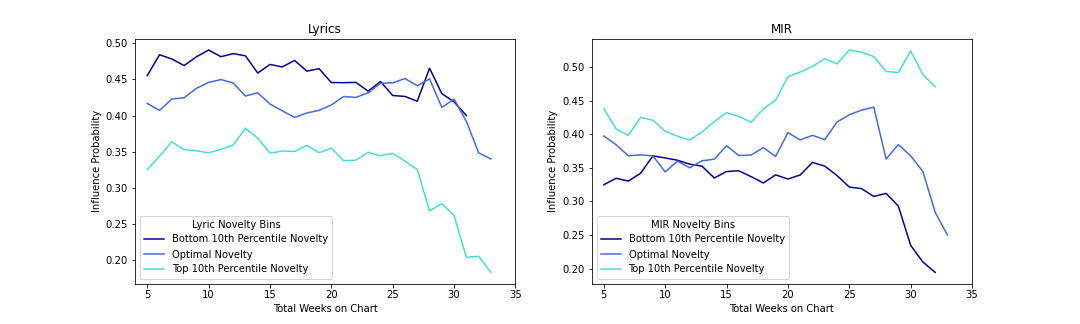}
\caption{Controlling for initial novelty, the influence probability increases at first with additional weeks on chart, then after a certain amount of time it peaks and begins decreasing. For lyric novelty, the peak occurs at roughly 10 weeks. For MIR novelty, the change in influence probability relative to time on chart varies depending on the initial novelty score. For higher initial novelty scores, we see the peak influence probability occur after a longer period of time on the chart. }
\label{fig:exposure effects}
\end{figure}

We found that, relative to the song's novelty score, the influence probability varied with increases in the amount of time the song spent on the chart. In Fig. \ref{fig:exposure effects} we see that regardless of the modality or the novelty grouping, influence probability initially increases with more time spent on chart, then at a certain point, peaks and starts to decrease. For lyric novelty, we see a peak occurs at the same time for all three novelty bins, at roughly 10-15 weeks. Beyond that, both the bottom 10th percentile and the top 10th percentile see a decrease, although the rate of decrease for influence probability appears to be more pronounced for the songs in the top 90th novelty percentile. For lyric novelty in the optimal range, we do see a second peak around 25-30 weeks, however beyond that the influence probability again drops. We see that regardless of the total amount of time on chart, higher lyric novelty scores are correlated with lower influence probability. For MIR novelty, we see in Fig.  \ref{fig:exposure effects} that the relationship between the influence probability and additional weeks on the chart varies depending on the song's novelty score. Songs in the bottom 10th percentile see influence probability peak at 10 weeks, and then consistently decrease with additional time on the chart. For songs in the optimal zone, however, there is a consistent increase in influence probability until roughly 26-7 weeks, at which point the influence probability quickly decreases. For songs in the top 90th percentile, we see a steeper increase in influence probability which lasts until roughly 30 weeks before hitting the peak and then decreasing. Here we also see that in contrast to the pattern observed for lyric novelty scores, higher MIR novelty scores are correlated with higher influence probability regardless of the number of weeks spent on the chart. We typically associate increased exposure with greater success, for both short term commercial success, and also for being influential within the creative space. It would seem intuitive that greater exposure would lead to a greater probability of exerting influence, as more people hear and become familiar with the cultural artifact in question. However these results suggest that while that is sometimes the case, there is also an optimal amount of exposure, which can vary depending on the novelty and modality of the attributes being considered.

\begin{figure}[ht]
\centering
\includegraphics[width=12cm]{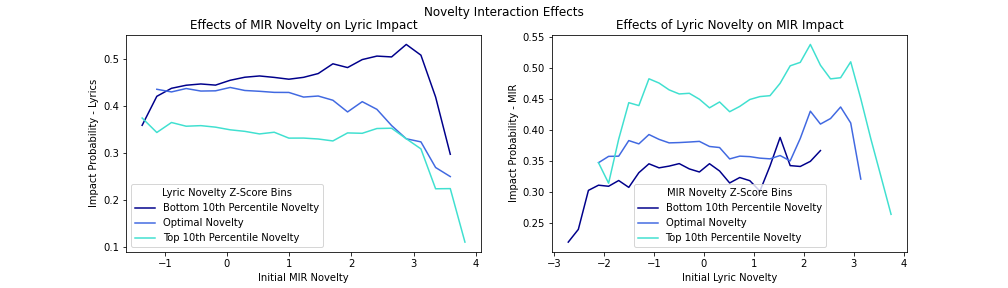}
\caption{Controlling for initial novelty, we see that different combinations of high and low MIR and lyric novelty scores have different associations with changes in influence probability.}
\label{fig:pull}
\end{figure}

Although we previously found no correlation between individual songs' initial lyric novelty and initial MIR novelty, we did find that there was a small but significant correlation (r=0.22, p$<$ 0.001) between the average change in lyric relative novelty and the average change in MIR relative novelty (see Fig. \ref{fig:pull}). Additionally, we observed that for songs with low lyric novelty scores, higher MIR novelty scores had a slight positive correlation with influence probability, while for songs with high MIR novelty scores, higher lyric novelty scores also had a slight positive correlation with influence probability.  A possible explanation is that these combinations of high and low novelty scores impact how memorable a song is, even if they are more likely to hurt the songs initial success probability. However, it is not clear why we observe these interaction patterns only for songs at the extreme ends of the novelty score ranges, and these findings highlight an avenue for further research into how the attributes of cultural artifacts impact their likelihood of influencing the larger cultural ecosystem. 

\subsection{Feature Variance Over Time}

\begin{figure}[ht]
\centering
\includegraphics[width=10cm]{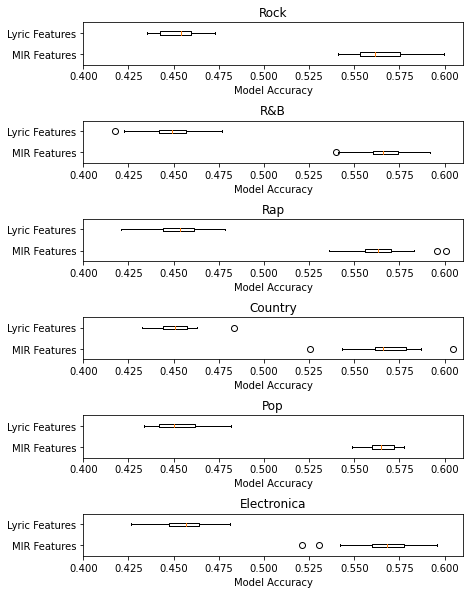}
\caption{Distributions of accuracies for a random forest classifier when trained on each genre's MIR and lyric features, respectively. The difference between each pair of distributions is statistically significant (p$<$ 0.001), with MIR features resulting in higher accuracy scores from cross-validation.}
\label{fig:RF}
\end{figure}

In order to delve into why the effect of exposure was so different for lyric novelty than for MIR novelty, we examine the differences in the amount of change over time for the average MIR features compared to the amount of change over time for the average lyric features. In Fig. \ref{fig:RF}, we compare the distribution of accuracy scores when using our random forest classifier to predict the temporal class of individual songs using the MIR features versus when using the lyric features. For each of the largest genres, we observed a statistically significant difference (p$<$ 0.001) between the predictive power of the two distributions, with the MIR features consistently resulting in more accurate predictions. Given that the overall accuracy scores are fairly low, with the model trained on lyric features ranging between 42\% and 48\% accuracy and the model trained on MIR features ranging between 52\% and 60\% accuracy, this indicates that there is still overlap in the feature distributions of different decades. However, since our analysis demonstrates that there is more predictive power in the MIR features, this tells us that the distributions of MIR features from different decades are more distinct from one another in the feature space than are the distributions of lyric vectors for different decades. Given that we also found that the average year-over-year change in relative novelty for MIR stays positive, indicating that the average distance between old and new songs in MIR feature space is always increasing, we can infer that within the individual genres, there is long-term directionality to the genre's movement through MIR feature space. In contrast, the average year-over-year change in relative novelty for lyrics is, for the most part, negative, which when taken in conjunction with the lower classification accuracy scores when using lyric features, suggests that there is not a significant amount of directionality to the movement occurring within the lyric feature space.

\section{Discussion}

By utilizing computational methods to analyze cultural data at scale, this research contributes to our understanding of the relationship between novelty impacts the dynamics of cultural change within the context of the larger cultural ecosystem. Our results highlight the ways in which we perceive and evaluate different degrees of novelty and differentiation. Although the high-level and aggregate nature of our data does not enable us to create a prediction model for identifying hit songs in their unique context, or causally attribute stylistic changes within a genre to the influence of specific songs, our results contribute to understanding overarching patterns  of novelty.

\subsection{Novelty and Music Cognition}
Our finding that there is no relationship between the lyric and MIR novelty scores of individual songs, and that the optimal novelty scores for each modality are also independent of one another is supported by previous work in music cognition, which finds evidence that music and lyrics are processed independently \cite{besson_singing_1998, rigoulot_early_2016}. This explains why the negative impact on success probability of an above optimal novelty score for one modality cannot be mitigated by the song having a below optimal score in the other modality. Our results do not provide any information which would allow us to evaluate the possibility of a causal relationship between the aforementioned music cognition data and the trends we have observed, however the observation of these connections between large scale phenomenon and cognitive processes that occur at the individual level suggest that this could be a productive area of interdisciplinary study. It is possible that research in the field of cognitive science could provide insights into cognitive perceptual processing, which could inform potential avenues of inquiry when investigating cultural trends and evolution.

\subsection{Novelty and Exposure Effects}

Although our data is not sufficient to investigate any possible causal relationship between initial novelty and influence probability, we observed that the relationship between a song's initial novelty score and its influence probability varies relative to how many weeks it has spent on the chart.  For songs with higher MIR novelty scores, we see that an increase in time on chart has a positive correlation with influence probability within the time range we analyzed. This potentially explains why the distribution of MIR novelty scores is more heavily right-tailed than the distribution of lyric novelty scores, as the positive impact of increased exposure may cause more variation and range in the MIR novelty scores that end up being successful enough to reach the Billboard Hot 100 chart. 

Previous work exploring the impact of repeated exposure to unfamiliar music and subsequent music preferences have found that this additional exposure increases the likelihood that the listeners will enjoy the music when they hear it again \cite{peretz_exposure_1998}. Additional research has also found that repeated exposure in the context of collective attention to news stories shared online, leads to novelty decay over time \cite{wu_novelty_2007}. It is possible, then, that for high-novelty songs, the increased exposure due to more time spent on the chart may decrease the perceived novelty of the song, leading audiences to experience it as being closer to the optimal level of differentiation. For low-novelty artifacts, however, this decrease in perceived novelty could make them seem too familiar, hence we see that the amount of exposure which was beneficial varied depending on the initial novelty scores. For lyric novelty, we saw that increased time on the chart correlated with a decrease in influence probability, regardless of whether the lyric novelty score was lower or higher than optimal. Given that we observed significantly less variance over time for lyric features than we did for MIR features, it is possible that even songs with very high lyric novelty scores are not distinct enough to benefit from higher levels of exposure.

For lyric novelty, we saw that increased time on the chart correlated with a decrease in influence probability, regardless of whether the lyric novelty score was lower or higher than optimal. Given that we observed significantly less variance over time for lyric features than we did for MIR features, it is possible that even songs with very high lyric novelty scores are not distinct enough to benefit from higher levels of exposure. This suggests that when analyzing  differences between cultural artifacts and their relationship to various metrics of success, it is important to draw a distinction between differentiation, which measures the amount of variation between the artifact and the other artifacts it is being compared to, and `true' novelty, which would consider the degree to which the artifact is introducing new material into the canon of its domain.

Our findings suggest that it is possible for an artifact to be `overexposed', at which point in time the perception of novelty drops below an optimal level. Our results indicate that the amount of exposure it takes for this to occur is going to vary depending on the initial novelty of the artifact. This is an important consideration when modeling and predicting the dissemination of creative ideas and products, both in theoretical research, and in the development of practical applications, such as recommender systems. Additionally, this highlights the importance of considering the potential effect of social influence on not only the initial popularity of cultural artifacts, but also the longer term evolution of the cultural ecosystem. In Salganik et al's study on social influence in cultural markets, different social behaviors led to different patterns of success within an artificial music market, demonstrating the significant impact of social influence on what songs become popular  \cite{salganik_experimental_2006}. While the effect of social influence was shown to be largely independent of the specific attributes of the individual songs, by impacting which songs become popular this will also impact the relative amount of exposure those songs will receive. As our analysis suggests that patterns of exposure may potentially impact the likelihood of an artifact exerting stylistic influence, suggests a possible mechanism by which the impact of social influence has a downstream effect on the stylistic evolution of the musical ecosystem.

\subsection{Implications for Recommender Systems}
Understanding the degree to which the defining characteristics of musical genres change over time has applications for music recommendation software. There are limitations to traditional approaches of using historical data in predictions \cite{jung_things_2020}, and in order to avoid static behavior it is important for us to be able to identify what are the indicators that can help predict a preference shift \cite{moore_taste_2013}. If we can determine the degree to which more novel outliers indicate the evolution of a subset of music, we can take that into account when tracking an individual's music preferences, and better predict what they might like as their taste evolves. This is especially relevant for increasing the efficiency of recommender systems incorporating exploratory algorithms, as it can inform more directed exploration, as well as the ideal degree of novelty to incorporate with each round of exploration \cite{xing_enhancing_2014}.

\subsection{Limitations}
It is important to note that there are many external factors that can affect whether or not a song become successful, which are unfortunately not captured in the scope of this data set. As a result, we cannot draw any inferences about direct causal relationships between novelty and the success metrics we examined. Additionally, our data is lacking important controls that could be correlated with novelty and influence, due to the currently unavailability of such comprehensive data.

As the Hot 100 data contains only a partial view of popular music, and its selection criteria has changed over time, future work could involve gathering a larger data set that would encompass a broader representation of modern music. For the purposes of this analysis, the Hot 100 songs served as a sample of the prevailing mainstream cultural trends in music. However, it is still a small sample of all the potential songs that could be included in the umbrella of modern popular music. Additionally when grouping music by genre, we must acknowledge that genre classifications are inherently subjective. Genre labeling for this data set came from Discogs.com, which provides crowd-sourced data for songs, so the groupings provided do not necessarily represent an objective ground truth \cite{lorince_wisdom_2015}. 

\section{Conclusion}
Utilizing MIR data to perform analysis at scale, we compared musical artifacts' relative novelty over time to identify consistent patterns in the dynamics of cultural change. Our results showed evidence for both optimal differentiation in successful songs, and the conditioning effect of prior artifacts on stylistic change. By bringing in findings from sociology, cognitive science, and musicology to provide further insight into the impact of novelty on modern music evolution, our research provides quantitative methods that will enable media systems to track this organic evolution in a more informed manner.


\begin{backmatter}

\section*{Competing interests}
  The authors declare that they have no competing interests.

\section*{Author's contributions}
   KO designed the study, ran all data analysis, and drafted the manuscript. E\'AH provided feedback on study design and revised and edited the manuscript. All authors read and approved the final manuscript. 

\section*{Acknowledgements}
  The authors would like to thank Miriam Alex for her assistance with data collection and cleaning, and would like to thank Michael Mauskapf and Noah Askin for sharing their unique data set.
  
\section*{Abbreviations}
MIR, Music Information Retrieval

\section*{Availability of data and materials}
The code that supports the findings of this study are available at: \url{https://github.com/LINK-NU/Billboard-Hot-100-Music-Novelty.} Data available on request.

\bibliographystyle{bmc-mathphys} 
\bibliography{references1.bib}    


\begin{thebibliography}{54}
\ifx \bisbn   \undefined \def \bisbn  #1{ISBN #1}\fi
\ifx \binits  \undefined \def \binits#1{#1}\fi
\ifx \bauthor  \undefined \def \bauthor#1{#1}\fi
\ifx \batitle  \undefined \def \batitle#1{#1}\fi
\ifx \bjtitle  \undefined \def \bjtitle#1{#1}\fi
\ifx \bvolume  \undefined \def \bvolume#1{\textbf{#1}}\fi
\ifx \byear  \undefined \def \byear#1{#1}\fi
\ifx \bissue  \undefined \def \bissue#1{#1}\fi
\ifx \bfpage  \undefined \def \bfpage#1{#1}\fi
\ifx \blpage  \undefined \def \blpage #1{#1}\fi
\ifx \burl  \undefined \def \burl#1{\textsf{#1}}\fi
\ifx \doiurl  \undefined \def \doiurl#1{\textsf{#1}}\fi
\ifx \betal  \undefined \def \betal{\textit{et al.}}\fi
\ifx \binstitute  \undefined \def \binstitute#1{#1}\fi
\ifx \binstitutionaled  \undefined \def \binstitutionaled#1{#1}\fi
\ifx \bctitle  \undefined \def \bctitle#1{#1}\fi
\ifx \beditor  \undefined \def \beditor#1{#1}\fi
\ifx \bpublisher  \undefined \def \bpublisher#1{#1}\fi
\ifx \bbtitle  \undefined \def \bbtitle#1{#1}\fi
\ifx \bedition  \undefined \def \bedition#1{#1}\fi
\ifx \bseriesno  \undefined \def \bseriesno#1{#1}\fi
\ifx \blocation  \undefined \def \blocation#1{#1}\fi
\ifx \bsertitle  \undefined \def \bsertitle#1{#1}\fi
\ifx \bsnm \undefined \def \bsnm#1{#1}\fi
\ifx \bsuffix \undefined \def \bsuffix#1{#1}\fi
\ifx \bparticle \undefined \def \bparticle#1{#1}\fi
\ifx \barticle \undefined \def \barticle#1{#1}\fi
\ifx \bconfdate \undefined \def \bconfdate #1{#1}\fi
\ifx \botherref \undefined \def \botherref #1{#1}\fi
\ifx \url \undefined \def \url#1{\textsf{#1}}\fi
\ifx \bchapter \undefined \def \bchapter#1{#1}\fi
\ifx \bbook \undefined \def \bbook#1{#1}\fi
\ifx \bcomment \undefined \def \bcomment#1{#1}\fi
\ifx \oauthor \undefined \def \oauthor#1{#1}\fi
\ifx \citeauthoryear \undefined \def \citeauthoryear#1{#1}\fi
\ifx \endbibitem  \undefined \def \endbibitem {}\fi
\ifx \bconflocation  \undefined \def \bconflocation#1{#1}\fi
\ifx \arxivurl  \undefined \def \arxivurl#1{\textsf{#1}}\fi
\csname PreBibitemsHook\endcsname

\bibitem{mauch_evolution_2015}
\begin{barticle}
\bauthor{\bsnm{Mauch}, \binits{M.}},
\bauthor{\bsnm{MacCallum}, \binits{R.M.}},
\bauthor{\bsnm{Levy}, \binits{M.}},
\bauthor{\bsnm{Leroi}, \binits{A.M.}}:
\batitle{The evolution of popular music: {USA} 1960–2010}.
\bjtitle{Royal Society Open Science}
\bvolume{2}(\bissue{5}),
\bfpage{150081}
(\byear{2015}).
doi:\doiurl{10.1098/rsos.150081}.
\bcomment{Publisher: Royal Society}
\end{barticle}
\endbibitem

\bibitem{weis_investigating_2019}
\begin{barticle}
\bauthor{\bsnm{Weiß}, \binits{C.}},
\bauthor{\bsnm{Mauch}, \binits{M.}},
\bauthor{\bsnm{Dixon}, \binits{S.}},
\bauthor{\bsnm{Müller}, \binits{M.}}:
\batitle{Investigating style evolution of {Western} classical music: {A}
  computational approach}.
\bjtitle{Musicae Scientiae}
\bvolume{23}(\bissue{4}),
\bfpage{486}--\blpage{507}
(\byear{2019}).
doi:\doiurl{10.1177/1029864918757595}
\end{barticle}
\endbibitem

\bibitem{serra_measuring_2012}
\begin{barticle}
\bauthor{\bsnm{Serrà}, \binits{J.}},
\bauthor{\bsnm{Corral}, \binits{A.}},
\bauthor{\bsnm{Boguñá}, \binits{M.}},
\bauthor{\bsnm{Haro}, \binits{M.}},
\bauthor{\bsnm{Arcos}, \binits{J.L.}}:
\batitle{Measuring the {Evolution} of {Contemporary} {Western} {Popular}
  {Music}}.
\bjtitle{Scientific Reports}
\bvolume{2}(\bissue{1}),
\bfpage{521}
(\byear{2012}).
doi:\doiurl{10.1038/srep00521}.
\bcomment{Number: 1 Publisher: Nature Publishing Group}
\end{barticle}
\endbibitem

\bibitem{bomin_evolution_2016}
\begin{barticle}
\bauthor{\bsnm{Bomin}, \binits{S.L.}},
\bauthor{\bsnm{Lecointre}, \binits{G.}},
\bauthor{\bsnm{Heyer}, \binits{E.}}:
\batitle{The {Evolution} of {Musical} {Diversity}: {The} {Key} {Role} of
  {Vertical} {Transmission}}.
\bjtitle{PLOS ONE}
\bvolume{11}(\bissue{3}),
\bfpage{0151570}
(\byear{2016}).
doi:\doiurl{10.1371/journal.pone.0151570}.
\bcomment{Publisher: Public Library of Science}
\end{barticle}
\endbibitem

\bibitem{prockup_modeling_2015}
\begin{bchapter}
\bauthor{\bsnm{Prockup}, \binits{M.}},
\bauthor{\bsnm{Ehmann}, \binits{A.F.}},
\bauthor{\bsnm{Gouyon}, \binits{F.}},
\bauthor{\bsnm{Schmidt}, \binits{E.M.}},
\bauthor{\bsnm{Celma}, \binits{O.}},
\bauthor{\bsnm{Kim}, \binits{Y.E.}}:
\bctitle{Modeling {Genre} with the {Music} {Genome} {Project}: {Comparing}
  {Human}-{Labeled} {Attributes} and {Audio} {Features}.}
In: \bbtitle{Proceedings of the 16th {ISMIR} {Conference}},
\bconflocation{Malaga, Spain},
p. \bfpage{7}
(\byear{2015})
\end{bchapter}
\endbibitem

\bibitem{klimek_fashion_2019}
\begin{barticle}
\bauthor{\bsnm{Klimek}, \binits{P.}},
\bauthor{\bsnm{Kreuzbauer}, \binits{R.}},
\bauthor{\bsnm{Thurner}, \binits{S.}}:
\batitle{Fashion and art cycles are driven by counter-dominance signals of
  elite competition: quantitative evidence from music styles}.
\bjtitle{Journal of The Royal Society Interface}
\bvolume{16}(\bissue{151}),
\bfpage{20180731}
(\byear{2019}).
doi:\doiurl{10.1098/rsif.2018.0731}
\end{barticle}
\endbibitem

\bibitem{magron_leveraging_2020}
\begin{botherref}
\oauthor{\bsnm{Magron}, \binits{P.}},
\oauthor{\bsnm{Févotte}, \binits{C.}}:
Leveraging the structure of musical preference in content-aware music
  recommendation.
CoRR
\textbf{abs/2010.10276}
(2020).
arXiv: 2010.10276
\end{botherref}
\endbibitem

\bibitem{moffat_evaluation_2015}
\begin{bchapter}
\bauthor{\bsnm{Moffat}, \binits{D.}},
\bauthor{\bsnm{Ronan}, \binits{D.}},
\bauthor{\bsnm{Reiss}, \binits{J.D.}}:
\bctitle{An {Evaluation} of {Audio} {Feature} {Extraction} {Toolboxes}}.
In: \bbtitle{Proc. of the 18th {Int}. {Conference} on {Digital} {Audio}
  {Effects} ({DAFx}-15)},
\bconflocation{Trondheim, Norway},
p. \bfpage{7}
(\byear{2015})
\end{bchapter}
\endbibitem

\bibitem{friberg_using_2014}
\begin{barticle}
\bauthor{\bsnm{Friberg}, \binits{A.}},
\bauthor{\bsnm{Schoonderwaldt}, \binits{E.}},
\bauthor{\bsnm{Hedblad}, \binits{A.}},
\bauthor{\bsnm{Fabiani}, \binits{M.}},
\bauthor{\bsnm{Elowsson}, \binits{A.}}:
\batitle{Using perceptually defined music features in music information
  retrieval}.
\bjtitle{arXiv:1403.7923 [cs]}
(\byear{2014}).
doi:\doiurl{10.48550/arXiv.1403.7923}
\end{barticle}
\endbibitem

\bibitem{bertin-mahieux_large-scale_2013}
\begin{botherref}
\oauthor{\bsnm{Bertin-Mahieux}, \binits{T.}}:
Large-{Scale} {Pattern} {Discovery} in {Music}.
PhD thesis,
Columbia University
(2013).
doi:\doiurl{10.7916/D8NC67CT}.
\url{https://doi.org/10.7916/D8NC67CT}
\end{botherref}
\endbibitem

\bibitem{lippens_comparison_2004}
\begin{bchapter}
\bauthor{\bsnm{Lippens}, \binits{S.}},
\bauthor{\bsnm{Martens}, \binits{J.P.}},
\bauthor{\bsnm{De~Mulder}, \binits{T.}}:
\bctitle{A comparison of human and automatic musical genre classification}.
In: \bbtitle{2004 {IEEE} {International} {Conference} on {Acoustics}, {Speech},
  and {Signal} {Processing}},
\bconflocation{Montreal, QC, Canada},
p.
(\byear{2004}).
doi:\doiurl{10.1109/ICASSP.2004.1326806}.
\bcomment{ISSN: 1520-6149}
\end{bchapter}
\endbibitem

\bibitem{interiano_musical_2018}
\begin{barticle}
\bauthor{\bsnm{Interiano}, \binits{M.}},
\bauthor{\bsnm{Kazemi}, \binits{K.}},
\bauthor{\bsnm{Wang}, \binits{L.}},
\bauthor{\bsnm{Yang}, \binits{J.}},
\bauthor{\bsnm{Yu}, \binits{Z.}},
\bauthor{\bsnm{Komarova}, \binits{N.L.}}:
\batitle{Musical trends and predictability of success in contemporary songs in
  and out of the top charts}.
\bjtitle{Royal Society Open Science}
\bvolume{5}(\bissue{5}),
\bfpage{171274}
(\byear{2018}).
doi:\doiurl{10.1098/rsos.171274}.
\bcomment{Publisher: Royal Society}
\end{barticle}
\endbibitem

\bibitem{mayerl_comparing_2020}
\begin{bchapter}
\bauthor{\bsnm{Mayerl}, \binits{M.}},
\bauthor{\bsnm{Votter}, \binits{M.}},
\bauthor{\bsnm{Zangerle}, \binits{M.M.E.}}:
\bctitle{Comparing {Lyrics} {Features} for {Genre} {Recognition}}.
In: \bbtitle{Proceedings of the 1st {Workshop} on {NLP} for {Music} and {Audio}
  ({NLP4MusA})},
pp. \bfpage{73}--\blpage{77}.
\bpublisher{Association for Computational Linguistics},
\blocation{Online}
(\byear{2020}).
\burl{https://www.aclweb.org/anthology/2020.nlp4musa-1.15.pdf}
\end{bchapter}
\endbibitem

\bibitem{mayer_combination_2008}
\begin{bchapter}
\bauthor{\bsnm{Mayer}, \binits{R.}},
\bauthor{\bsnm{Neumayer}, \binits{R.}},
\bauthor{\bsnm{Rauber}, \binits{A.}}:
\bctitle{Combination of audio and lyrics features for genre classification in
  digital audio collections}.
In: \bbtitle{Proceedings of the 16th {ACM} International Conference on
  {Multimedia}}.
\bsertitle{{MM} '08},
pp. \bfpage{159}--\blpage{168}.
\bpublisher{Association for Computing Machinery},
\blocation{New York, NY, USA}
(\byear{2008}).
doi:\doiurl{10.1145/1459359.1459382}.
\burl{https://doi.org/10.1145/1459359.1459382}
\end{bchapter}
\endbibitem

\bibitem{hu_lyric_2009}
\begin{bchapter}
\bauthor{\bsnm{Hu}, \binits{X.}},
\bauthor{\bsnm{Downie}, \binits{J.S.}},
\bauthor{\bsnm{Ehmann}, \binits{A.F.}}:
\bctitle{Lyric {Text} {Mining} in {Music} {Mood} {Classification}}.
In: \bbtitle{Proceedings of the 10th {International} {Society} for {Music}
  {Information} {Retrieval} {Conference}},
\bconflocation{Kobe International Conference Center, Kobe, Japan},
p. \bfpage{6}
(\byear{2009})
\end{bchapter}
\endbibitem

\bibitem{hu_improving_2010}
\begin{bchapter}
\bauthor{\bsnm{Hu}, \binits{X.}},
\bauthor{\bsnm{Downie}, \binits{J.S.}}:
\bctitle{Improving mood classification in music digital libraries by combining
  lyrics and audio}.
In: \bbtitle{Proceedings of the 10th Annual Joint Conference on {Digital}
  Libraries}.
\bsertitle{{JCDL} '10},
pp. \bfpage{159}--\blpage{168}.
\bpublisher{Association for Computing Machinery},
\blocation{New York, NY, USA}
(\byear{2010}).
doi:\doiurl{10.1145/1816123.1816146}.
\burl{https://doi.org/10.1145/1816123.1816146}
\end{bchapter}
\endbibitem

\bibitem{mcvicar_lyric_2021}
\begin{bchapter}
\bauthor{\bsnm{McVicar}, \binits{M.}},
\bauthor{\bsnm{Giorgi}, \binits{B.D.}},
\bauthor{\bsnm{Dundar}, \binits{B.}},
\bauthor{\bsnm{Mauch}, \binits{M.}}:
\bctitle{Lyric document embeddings for music tagging}.
In: \bbtitle{Proc. of the 15th {International} {Symposium} On {CMMR}},
\bconflocation{Online},
p. \bfpage{10}
(\byear{2021})
\end{bchapter}
\endbibitem

\bibitem{mikolov_efficient_2013}
\begin{botherref}
\oauthor{\bsnm{Mikolov}, \binits{T.}},
\oauthor{\bsnm{Chen}, \binits{K.}},
\oauthor{\bsnm{Corrado}, \binits{G.}},
\oauthor{\bsnm{Dean}, \binits{J.}}:
Efficient {Estimation} of {Word} {Representations} in {Vector} {Space}.
arXiv:1301.3781 [cs]
(2013).
arXiv: 1301.3781
\end{botherref}
\endbibitem

\bibitem{le_distributed_2014}
\begin{botherref}
\oauthor{\bsnm{Le}, \binits{Q.V.}},
\oauthor{\bsnm{Mikolov}, \binits{T.}}:
Distributed {Representations} of {Sentences} and {Documents}.
arXiv:1405.4053 [cs]
(2014).
arXiv: 1405.4053
\end{botherref}
\endbibitem

\bibitem{whalen_patent_2020}
\begin{barticle}
\bauthor{\bsnm{Whalen}, \binits{R.}},
\bauthor{\bsnm{Lungeanu}, \binits{A.}},
\bauthor{\bsnm{DeChurch}, \binits{L.}},
\bauthor{\bsnm{Contractor}, \binits{N.}}:
\batitle{Patent {Similarity} {Data} and {Innovation} {Metrics}}.
\bjtitle{Journal of Empirical Legal Studies}
\bvolume{17}(\bissue{3}),
\bfpage{615}--\blpage{639}
(\byear{2020}).
doi:\doiurl{10.1111/jels.12261}.
\bcomment{eprint: https://onlinelibrary.wiley.com/doi/pdf/10.1111/jels.12261}
\end{barticle}
\endbibitem

\bibitem{askin_what_2017}
\begin{barticle}
\bauthor{\bsnm{Askin}, \binits{N.}},
\bauthor{\bsnm{Mauskapf}, \binits{M.}}:
\batitle{What {Makes} {Popular} {Culture} {Popular}? {Product} {Features} and
  {Optimal} {Differentiation} in {Music}}.
\bjtitle{American Sociological Review}
\bvolume{82}(\bissue{5}),
\bfpage{910}--\blpage{944}
(\byear{2017}).
doi:\doiurl{10.1177/0003122417728662}
\end{barticle}
\endbibitem

\bibitem{askin_cultural_2014}
\begin{bchapter}
\bauthor{\bsnm{Askin}, \binits{N.}},
\bauthor{\bsnm{Mauskapf}, \binits{M.}}:
\bctitle{Cultural {Attributes} and their {Influence} on {Consumption}
  {Patterns} in {Popular} {Music}}.
In: \beditor{\bsnm{Aiello}, \binits{L.M.}},
\beditor{\bsnm{McFarland}, \binits{D.}} (eds.)
\bbtitle{(eds) {Social} {Informatics}. {SocInfo} 2014}.
\bsertitle{Lecture {Notes} in {Computer} {Science}},
vol. \bseriesno{8851},
pp. \bfpage{508}--\blpage{530}.
\bpublisher{Springer},
\blocation{Cham}
(\byear{2014}).
\burl{https://doi.org/10.1007/978-3-319-13734-6$\_$36}
\end{bchapter}
\endbibitem

\bibitem{berger_are_2018}
\begin{barticle}
\bauthor{\bsnm{Berger}, \binits{J.}},
\bauthor{\bsnm{Packard}, \binits{G.}}:
\batitle{Are {Atypical} {Things} {More} {Popular}?}
\bjtitle{Psychological Science}
\bvolume{29}(\bissue{7}),
\bfpage{1178}--\blpage{1184}
(\byear{2018}).
doi:\doiurl{10.1177/0956797618759465}.
\bcomment{Publisher: SAGE Publications Inc}
\end{barticle}
\endbibitem

\bibitem{anderson_algorithmic_2020}
\begin{bchapter}
\bauthor{\bsnm{Anderson}, \binits{A.}},
\bauthor{\bsnm{Maystre}, \binits{L.}},
\bauthor{\bsnm{Anderson}, \binits{I.}},
\bauthor{\bsnm{Mehrotra}, \binits{R.}},
\bauthor{\bsnm{Lalmas}, \binits{M.}}:
\bctitle{Algorithmic {Effects} on the {Diversity} of {Consumption} on
  {Spotify}}.
In: \bbtitle{Proceedings of {The} {Web} {Conference} 2020},
pp. \bfpage{2155}--\blpage{2165}.
\bpublisher{Association for Computing Machinery},
\blocation{New York, NY, USA}
(\byear{2020}).
\burl{https://doi.org/10.1145/3366423.3380281}
\end{bchapter}
\endbibitem

\bibitem{laurier_multimodal_2008}
\begin{bchapter}
\bauthor{\bsnm{Laurier}, \binits{C.}},
\bauthor{\bsnm{Grivolla}, \binits{J.}},
\bauthor{\bsnm{Herrera}, \binits{P.}}:
\bctitle{Multimodal {Music} {Mood} {Classification} {Using} {Audio} and
  {Lyrics}}.
In: \bbtitle{2008 {Seventh} {International} {Conference} on {Machine}
  {Learning} and {Applications}},
pp. \bfpage{688}--\blpage{693}
(\byear{2008}).
doi:\doiurl{10.1109/ICMLA.2008.96}
\end{bchapter}
\endbibitem

\bibitem{amati_integration_2007}
\begin{bchapter}
\bauthor{\bsnm{Neumayer}, \binits{R.}},
\bauthor{\bsnm{Rauber}, \binits{A.}}:
\bctitle{Integration of {Text} and {Audio} {Features} for {Genre}
  {Classification} in {Music} {Information} {Retrieval}}.
In: \beditor{\bsnm{Amati}, \binits{G.}},
\beditor{\bsnm{Carpineto}, \binits{C.}},
\beditor{\bsnm{Romano}, \binits{G.}} (eds.)
\bbtitle{Advances in {Information} {Retrieval}}.
\bsertitle{Lecture {Notes} in {Computer} {Science}},
vol. \bseriesno{4425},
pp. \bfpage{724}--\blpage{727}.
\bpublisher{Springer},
\blocation{Berlin, Heidelberg}
(\byear{2007}).
\burl{http://link.springer.com/10.1007/978-3-540-71496-5$\_$78}
\end{bchapter}
\endbibitem

\bibitem{saleh_toward_2016}
\begin{barticle}
\bauthor{\bsnm{Saleh}, \binits{B.}},
\bauthor{\bsnm{Abe}, \binits{K.}},
\bauthor{\bsnm{Arora}, \binits{R.S.}},
\bauthor{\bsnm{Elgammal}, \binits{A.}}:
\batitle{Toward automated discovery of artistic influence}.
\bjtitle{Multimedia Tools and Applications}
\bvolume{75}(\bissue{7}),
\bfpage{3565}--\blpage{3591}
(\byear{2016}).
doi:\doiurl{10.1007/s11042-014-2193-x}
\end{barticle}
\endbibitem

\bibitem{uzzi_atypical_2013}
\begin{barticle}
\bauthor{\bsnm{Uzzi}, \binits{B.}},
\bauthor{\bsnm{Mukherjee}, \binits{S.}},
\bauthor{\bsnm{Stringer}, \binits{M.}},
\bauthor{\bsnm{Jones}, \binits{B.}}:
\batitle{Atypical {Combinations} and {Scientific} {Impact}}.
\bjtitle{Science}
\bvolume{342}(\bissue{6157}),
\bfpage{468}--\blpage{472}
(\byear{2013}).
doi:\doiurl{10.1126/science.1240474}
\end{barticle}
\endbibitem

\bibitem{li_analyzing_2022}
\begin{bchapter}
\bauthor{\bsnm{Li}, \binits{Y.}},
\bauthor{\bsnm{Zhang}, \binits{Y.}},
\bauthor{\bsnm{Capra}, \binits{R.}}:
\bctitle{Analyzing information resources that support the creative process}.
In: \bbtitle{{ACM} {SIGIR} {Conference} on {Human} {Information} {Interaction}
  And {Retrieval}}.
\bsertitle{{CHIIR} '22},
pp. \bfpage{180}--\blpage{190}.
\bpublisher{Association for Computing Machinery},
\blocation{New York, NY, USA}
(\byear{2022}).
doi:\doiurl{10.1145/3498366.3505817}.
\burl{https://doi.org/10.1145/3498366.3505817}
Accessed 2022-04-01
\end{bchapter}
\endbibitem

\bibitem{liu_hot_2018}
\begin{barticle}
\bauthor{\bsnm{Liu}, \binits{L.}},
\bauthor{\bsnm{Wang}, \binits{Y.}},
\bauthor{\bsnm{Sinatra}, \binits{R.}},
\bauthor{\bsnm{Giles}, \binits{C.L.}},
\bauthor{\bsnm{Song}, \binits{C.}},
\bauthor{\bsnm{Wang}, \binits{D.}}:
\batitle{Hot streaks in artistic, cultural, and scientific careers}.
\bjtitle{Nature}
\bvolume{559}(\bissue{7714}),
\bfpage{396}--\blpage{399}
(\byear{2018}).
doi:\doiurl{10.1038/s41586-018-0315-8}.
\bcomment{Number: 7714 Publisher: Nature Publishing Group}.
Accessed 2020-10-14
\end{barticle}
\endbibitem

\bibitem{shin_scientific_2022}
\begin{barticle}
\bauthor{\bsnm{Shin}, \binits{H.}},
\bauthor{\bsnm{Kim}, \binits{K.}},
\bauthor{\bsnm{Kogler}, \binits{D.F.}}:
\batitle{Scientific collaboration, research funding, and novelty in scientific
  knowledge}.
\bjtitle{PLOS ONE}
\bvolume{17}(\bissue{7}),
\bfpage{0271678}
(\byear{2022}).
doi:\doiurl{10.1371/journal.pone.0271678}.
\bcomment{Publisher: Public Library of Science}.
Accessed 2022-08-02
\end{barticle}
\endbibitem

\bibitem{shi_weaving_2015}
\begin{barticle}
\bauthor{\bsnm{Shi}, \binits{F.}},
\bauthor{\bsnm{Foster}, \binits{J.G.}},
\bauthor{\bsnm{Evans}, \binits{J.A.}}:
\batitle{Weaving the fabric of science: {Dynamic} network models of science's
  unfolding structure}.
\bjtitle{Social Networks}
\bvolume{43},
\bfpage{73}--\blpage{85}
(\byear{2015}).
doi:\doiurl{10.1016/j.socnet.2015.02.006}.
Accessed 2021-01-26
\end{barticle}
\endbibitem

\bibitem{miles_what_2021}
\begin{botherref}
\oauthor{\bsnm{Miles}, \binits{S.A.}},
\oauthor{\bsnm{Rosen}, \binits{D.S.}},
\oauthor{\bsnm{Barry}, \binits{S.}},
\oauthor{\bsnm{Grunberg}, \binits{D.}},
\oauthor{\bsnm{Grzywacz}, \binits{N.}}:
What to {Expect} {When} the {Unexpected} {Becomes} {Expected}: {Harmonic}
  {Surprise} and {Preference} {Over} {Time} in {Popular} {Music}.
Frontiers in Human Neuroscience
\textbf{15}
(2021).
Accessed 2022-05-06
\end{botherref}
\endbibitem

\bibitem{sreenivasan_quantitative_2013}
\begin{barticle}
\bauthor{\bsnm{Sreenivasan}, \binits{S.}}:
\batitle{Quantitative analysis of the evolution of novelty in cinema through
  crowdsourced keywords}.
\bjtitle{Scientific Reports}
\bvolume{3}(\bissue{1}),
\bfpage{2758}
(\byear{2013}).
doi:\doiurl{10.1038/srep02758}.
\bcomment{Number: 1 Publisher: Nature Publishing Group}
\end{barticle}
\endbibitem

\bibitem{jing_sameness_2019}
\begin{botherref}
\oauthor{\bsnm{Jing}, \binits{E.}},
\oauthor{\bsnm{DeDeo}, \binits{S.}},
\oauthor{\bsnm{Ahn}, \binits{Y.-Y.}}:
Sameness {Attracts}, {Novelty} {Disturbs}, but {Outliers} {Flourish} in
  {Fanfiction} {Online}.
arXiv:1904.07741 [cs]
(2019).
arXiv: 1904.07741
\end{botherref}
\endbibitem

\bibitem{park_novelty_2020}
\begin{barticle}
\bauthor{\bsnm{Park}, \binits{D.}},
\bauthor{\bsnm{Nam}, \binits{J.}},
\bauthor{\bsnm{Park}, \binits{J.}}:
\batitle{Novelty and influence of creative works, and quantifying patterns of
  advances based on probabilistic references networks}.
\bjtitle{EPJ Data Science}
\bvolume{9}(\bissue{1}),
\bfpage{1}--\blpage{15}
(\byear{2020}).
doi:\doiurl{10.1140/epjds/s13688-019-0214-8}.
\bcomment{Number: 1 Publisher: SpringerOpen}.
Accessed 2020-10-12
\end{barticle}
\endbibitem

\bibitem{liu_pandemics_2022}
\begin{barticle}
\bauthor{\bsnm{Liu}, \binits{M.}},
\bauthor{\bsnm{Bu}, \binits{Y.}},
\bauthor{\bsnm{Chen}, \binits{C.}},
\bauthor{\bsnm{Xu}, \binits{J.}},
\bauthor{\bsnm{Li}, \binits{D.}},
\bauthor{\bsnm{Leng}, \binits{Y.}},
\bauthor{\bsnm{Freeman}, \binits{R.B.}},
\bauthor{\bsnm{Meyer}, \binits{E.T.}},
\bauthor{\bsnm{Yoon}, \binits{W.}},
\bauthor{\bsnm{Sung}, \binits{M.}},
\bauthor{\bsnm{Jeong}, \binits{M.}},
\bauthor{\bsnm{Lee}, \binits{J.}},
\bauthor{\bsnm{Kang}, \binits{J.}},
\bauthor{\bsnm{Min}, \binits{C.}},
\bauthor{\bsnm{Song}, \binits{M.}},
\bauthor{\bsnm{Zhai}, \binits{Y.}},
\bauthor{\bsnm{Ding}, \binits{Y.}}:
\batitle{Pandemics are catalysts of scientific novelty: {Evidence} from
  {COVID}-19}.
\bjtitle{Journal of the Association for Information Science and Technology}
\bvolume{73}(\bissue{8}),
\bfpage{1065}--\blpage{1078}
(\byear{2022}).
doi:\doiurl{10.1002/asi.24612}.
\bcomment{\_eprint: https://onlinelibrary.wiley.com/doi/pdf/10.1002/asi.24612}
\end{barticle}
\endbibitem

\bibitem{cheng_exploring_2020}
\begin{botherref}
\oauthor{\bsnm{Cheng}, \binits{D.}},
\oauthor{\bsnm{Joachims}, \binits{T.}},
\oauthor{\bsnm{Turnbull}, \binits{D.}}:
Exploring {Acoustic} {Similartiy} for {Novel} {Music} {Recommendation},
7
(2020)
\end{botherref}
\endbibitem

\bibitem{zangerle_hit_2019}
\begin{botherref}
\oauthor{\bsnm{Zangerle}, \binits{E.}},
\oauthor{\bsnm{Huber}, \binits{R.}},
\oauthor{\bsnm{Vötter}, \binits{M.}},
\oauthor{\bsnm{Yang}, \binits{Y.H.}}:
Hit song prediction: {Leveraging} low- and high-level audio features.
Proceedings of the 20th International Society for Music Information Retrieval
  Conference, ISMIR 2019,
319--326
(2019).
doi:\doiurl{10.5281/zenodo.3258042}.
Accessed 2020-11-01
\end{botherref}
\endbibitem

\bibitem{moore_taste_2013}
\begin{botherref}
\oauthor{\bsnm{Moore}, \binits{J.L.}},
\oauthor{\bsnm{Chen}, \binits{S.}},
\oauthor{\bsnm{Joachims}, \binits{T.}},
\oauthor{\bsnm{Turnbull}, \binits{D.}}:
Taste {Over} {Time}: {The} {Temporal} {Dynamics} of {User} {Preferences}.
Proceedings of the 14th International Society for Music Information Retrieval
  Conference, ISMIR 2013,
6
(2013)
\end{botherref}
\endbibitem

\bibitem{berlyne_novelty_1970}
\begin{barticle}
\bauthor{\bsnm{Berlyne}, \binits{D.E.}}:
\batitle{Novelty, complexity, and hedonic value}.
\bjtitle{Perception \& Psychophysics}
\bvolume{8}(\bissue{5}),
\bfpage{279}--\blpage{286}
(\byear{1970}).
doi:\doiurl{10.3758/BF03212593}.
Accessed 2021-03-19
\end{barticle}
\endbibitem

\bibitem{chmiel_back_2017}
\begin{barticle}
\bauthor{\bsnm{Chmiel}, \binits{A.}},
\bauthor{\bsnm{Schubert}, \binits{E.}}:
\batitle{Back to the inverted-{U} for music preference: {A} review of the
  literature}.
\bjtitle{Psychology of Music}
\bvolume{45}(\bissue{6}),
\bfpage{886}--\blpage{909}
(\byear{2017}).
doi:\doiurl{10.1177/0305735617697507}
\end{barticle}
\endbibitem

\bibitem{chai_breakthrough_2019}
\begin{barticle}
\bauthor{\bsnm{Chai}, \binits{S.}},
\bauthor{\bsnm{Menon}, \binits{A.}}:
\batitle{Breakthrough recognition: {Bias} against novelty and competition for
  attention}.
\bjtitle{Research Policy}
\bvolume{48}(\bissue{3}),
\bfpage{733}--\blpage{747}
(\byear{2019}).
doi:\doiurl{10.1016/j.respol.2018.11.006}
\end{barticle}
\endbibitem

\bibitem{wang_bias_2017}
\begin{barticle}
\bauthor{\bsnm{Wang}, \binits{J.}},
\bauthor{\bsnm{Veugelers}, \binits{R.}},
\bauthor{\bsnm{Stephan}, \binits{P.}}:
\batitle{Bias against novelty in science: {A} cautionary tale for users of
  bibliometric indicators}.
\bjtitle{Research Policy}
\bvolume{46}(\bissue{8}),
\bfpage{1416}--\blpage{1436}
(\byear{2017}).
\bcomment{Publisher: Elsevier}
\end{barticle}
\endbibitem

\bibitem{radim_rehurek_gensim_2010}
\begin{bchapter}
\bauthor{\bsnm{{Radim Řehůřek}}},
\bauthor{\bsnm{Sojka}, \binits{P.}}:
\bctitle{Gensim: topic modelling for humans}.
In: \bbtitle{Proceedings of the {LREC} 2010 {Workshop} on {New} {Challenges}
  for {NLP} {Frameworks}},
pp. \bfpage{45}--\blpage{50}.
\bpublisher{ELRA},
\blocation{Valletta, Malta}
(\byear{2010}).
\burl{https://radimrehurek.com/gensim/models/doc2vec.html}
\end{bchapter}
\endbibitem

\bibitem{pedregosa_scikit-learn_2011}
\begin{barticle}
\bauthor{\bsnm{Pedregosa}, \binits{F.}},
\bauthor{\bsnm{Varoquaux}, \binits{G.}},
\bauthor{\bsnm{Gramfort}, \binits{A.}},
\bauthor{\bsnm{Michel}, \binits{V.}},
\bauthor{\bsnm{Thirion}, \binits{B.}},
\bauthor{\bsnm{Grisel}, \binits{O.}},
\bauthor{\bsnm{Blondel}, \binits{M.}},
\bauthor{\bsnm{Prettenhofer}, \binits{P.}},
\bauthor{\bsnm{Weiss}, \binits{R.}},
\bauthor{\bsnm{Dubourg}, \binits{V.}},
\bauthor{\bsnm{Vanderplas}, \binits{J.}},
\bauthor{\bsnm{Passos}, \binits{A.}},
\bauthor{\bsnm{Cournapeau}, \binits{D.}},
\bauthor{\bsnm{Brucher}, \binits{M.}},
\bauthor{\bsnm{Perrot}, \binits{M.}},
\bauthor{\bsnm{Duchesnay}, \binits{E.}}:
\batitle{Scikit-learn: {Machine} {Learning} in {Python}}.
\bjtitle{Journal of Machine Learning Research}
\bvolume{12}(\bissue{85}),
\bfpage{2825}--\blpage{2830}
(\byear{2011})
\end{barticle}
\endbibitem

\bibitem{besson_singing_1998}
\begin{barticle}
\bauthor{\bsnm{Besson}, \binits{M.}},
\bauthor{\bsnm{Faïta}, \binits{F.}},
\bauthor{\bsnm{Peretz}, \binits{I.}},
\bauthor{\bsnm{Bonnel}, \binits{A.-M.}},
\bauthor{\bsnm{Requin}, \binits{J.}}:
\batitle{Singing in the {Brain}: {Independence} of {Lyrics} and {Tunes}}.
\bjtitle{Psychological Science}
\bvolume{9}(\bissue{6}),
\bfpage{494}--\blpage{498}
(\byear{1998}).
doi:\doiurl{10.1111/1467-9280.00091}.
\bcomment{Publisher: SAGE Publications Inc}
\end{barticle}
\endbibitem

\bibitem{rigoulot_early_2016}
\begin{barticle}
\bauthor{\bsnm{Rigoulot}, \binits{S.}},
\bauthor{\bsnm{Armony}, \binits{J.L.}}:
\batitle{Early selectivity for vocal and musical sounds: electrophysiological
  evidence from an adaptation paradigm}.
\bjtitle{European Journal of Neuroscience}
\bvolume{44}(\bissue{10}),
\bfpage{2786}--\blpage{2794}
(\byear{2016}).
doi:\doiurl{10.1111/ejn.13391}.
\bcomment{\_eprint: https://onlinelibrary.wiley.com/doi/pdf/10.1111/ejn.13391}
\end{barticle}
\endbibitem

\bibitem{peretz_exposure_1998}
\begin{barticle}
\bauthor{\bsnm{Peretz}, \binits{I.}},
\bauthor{\bsnm{Gaudreau}, \binits{D.}},
\bauthor{\bsnm{Bonnel}, \binits{A.-M.}}:
\batitle{Exposure effects on music preference and recognition}.
\bjtitle{Memory \& Cognition}
\bvolume{26}(\bissue{5}),
\bfpage{884}--\blpage{902}
(\byear{1998}).
doi:\doiurl{10.3758/BF03201171}
\end{barticle}
\endbibitem

\bibitem{wu_novelty_2007}
\begin{barticle}
\bauthor{\bsnm{Wu}, \binits{F.}},
\bauthor{\bsnm{Huberman}, \binits{B.A.}}:
\batitle{Novelty and collective attention}.
\bjtitle{Proceedings of the National Academy of Sciences}
\bvolume{104}(\bissue{45}),
\bfpage{17599}--\blpage{17601}
(\byear{2007}).
doi:\doiurl{10.1073/pnas.0704916104}.
\bcomment{Publisher: Proceedings of the National Academy of Sciences}
\end{barticle}
\endbibitem

\bibitem{salganik_experimental_2006}
\begin{barticle}
\bauthor{\bsnm{Salganik}, \binits{M.J.}},
\bauthor{\bsnm{Dodds}, \binits{P.S.}},
\bauthor{\bsnm{Watts}, \binits{D.J.}}:
\batitle{Experimental {Study} of {Inequality} and {Unpredictability} in an
  {Artificial} {Cultural} {Market}}.
\bjtitle{Science}
\bvolume{311}(\bissue{5762}),
\bfpage{854}--\blpage{856}
(\byear{2006}).
doi:\doiurl{10.1126/science.1121066}.
\bcomment{Publisher: American Association for the Advancement of Science
  Section: Report}
\end{barticle}
\endbibitem

\bibitem{jung_things_2020}
\begin{bchapter}
\bauthor{\bsnm{Jung}, \binits{S.-G.}},
\bauthor{\bsnm{Salminen}, \binits{J.}},
\bauthor{\bsnm{Chowdhury}, \binits{S.A.}},
\bauthor{\bsnm{Ramirez~Robillos}, \binits{D.}},
\bauthor{\bsnm{Jansen}, \binits{B.J.}}:
\bctitle{Things {Change}: {Comparing} {Results} {Using} {Historical} {Data} and
  {User} {Testing} for {Evaluating} a {Recommendation} {Task}}.
In: \bbtitle{Extended {Abstracts} of the 2020 {CHI} {Conference} on {Human}
  {Factors} in {Computing} {Systems}}.
\bsertitle{{CHI} {EA} '20},
pp. \bfpage{1}--\blpage{7}.
\bpublisher{Association for Computing Machinery},
\blocation{New York, NY, USA}
(\byear{2020}).
doi:\doiurl{10.1145/3334480.3382945}.
\burl{https://doi.org/10.1145/3334480.3382945}
\end{bchapter}
\endbibitem

\bibitem{xing_enhancing_2014}
\begin{bchapter}
\bauthor{\bsnm{Xing}, \binits{Z.}},
\bauthor{\bsnm{Wang}, \binits{X.}},
\bauthor{\bsnm{Wang}, \binits{Y.}}:
\bctitle{Enhancing {Collaborative} {Filtering} {Music} {Recommendation} by
  {Balancing} {Exploration} and {Exploitation}}.
In: \bbtitle{Proceedings of the 15th {International} {Society} for {Music}
  {Information} {Retrieval} {Conference}},
\bconflocation{Taipei, Taiwan}
(\byear{2014})
\end{bchapter}
\endbibitem

\bibitem{lorince_wisdom_2015}
\begin{botherref}
\oauthor{\bsnm{Lorince}, \binits{J.}},
\oauthor{\bsnm{Zorowitz}, \binits{S.}},
\oauthor{\bsnm{Murdock}, \binits{J.}},
\oauthor{\bsnm{Todd}, \binits{P.M.}}:
The {Wisdom} of the {Few}? “{Supertaggers}” in {Collaborative} {Tagging}
  {Systems}.
The Journal of Web Science
\textbf{1}
(2015).
doi:\doiurl{10.1561/106.00000002}
\end{botherref}
\endbibitem

\end{thebibliography}

\newcommand{\BMCxmlcomment}[1]{}

\BMCxmlcomment{

<refgrp>

<bibl id="B1">
  <title><p>The evolution of popular music: {USA} 1960–2010</p></title>
  <aug>
    <au><snm>Mauch</snm><fnm>M</fnm></au>
    <au><snm>MacCallum</snm><fnm>RM</fnm></au>
    <au><snm>Levy</snm><fnm>M</fnm></au>
    <au><snm>Leroi</snm><fnm>AM</fnm></au>
  </aug>
  <source>Royal Society Open Science</source>
  <pubdate>2015</pubdate>
  <volume>2</volume>
  <issue>5</issue>
  <fpage>150081</fpage>
  <url>https://royalsocietypublishing.org/doi/10.1098/rsos.150081</url>
  <note>Publisher: Royal Society</note>
</bibl>

<bibl id="B2">
  <title><p>Investigating style evolution of {Western} classical music: {A}
  computational approach</p></title>
  <aug>
    <au><snm>Weiß</snm><fnm>C</fnm></au>
    <au><snm>Mauch</snm><fnm>M</fnm></au>
    <au><snm>Dixon</snm><fnm>S</fnm></au>
    <au><snm>Müller</snm><fnm>M</fnm></au>
  </aug>
  <source>Musicae Scientiae</source>
  <pubdate>2019</pubdate>
  <volume>23</volume>
  <issue>4</issue>
  <fpage>486</fpage>
  <lpage>-507</lpage>
  <url>http://journals.sagepub.com/doi/10.1177/1029864918757595</url>
</bibl>

<bibl id="B3">
  <title><p>Measuring the {Evolution} of {Contemporary} {Western} {Popular}
  {Music}</p></title>
  <aug>
    <au><snm>Serrà</snm><fnm>J</fnm></au>
    <au><snm>Corral</snm><fnm>A</fnm></au>
    <au><snm>Boguñá</snm><fnm>M</fnm></au>
    <au><snm>Haro</snm><fnm>M</fnm></au>
    <au><snm>Arcos</snm><fnm>JL</fnm></au>
  </aug>
  <source>Scientific Reports</source>
  <pubdate>2012</pubdate>
  <volume>2</volume>
  <issue>1</issue>
  <fpage>521</fpage>
  <url>http://www.nature.com/articles/srep00521</url>
  <note>Number: 1 Publisher: Nature Publishing Group</note>
</bibl>

<bibl id="B4">
  <title><p>The {Evolution} of {Musical} {Diversity}: {The} {Key} {Role} of
  {Vertical} {Transmission}</p></title>
  <aug>
    <au><snm>Bomin</snm><fnm>SL</fnm></au>
    <au><snm>Lecointre</snm><fnm>G</fnm></au>
    <au><snm>Heyer</snm><fnm>E</fnm></au>
  </aug>
  <source>PLOS ONE</source>
  <pubdate>2016</pubdate>
  <volume>11</volume>
  <issue>3</issue>
  <fpage>e0151570</fpage>
  <url>https://journals.plos.org/plosone/article?id=10.1371/journal.pone.0151570</url>
  <note>Publisher: Public Library of Science</note>
</bibl>

<bibl id="B5">
  <title><p>Modeling {Genre} with the {Music} {Genome} {Project}: {Comparing}
  {Human}-{Labeled} {Attributes} and {Audio} {Features}.</p></title>
  <aug>
    <au><snm>Prockup</snm><fnm>M</fnm></au>
    <au><snm>Ehmann</snm><fnm>AF</fnm></au>
    <au><snm>Gouyon</snm><fnm>F</fnm></au>
    <au><snm>Schmidt</snm><fnm>EM</fnm></au>
    <au><snm>Celma</snm><fnm>O</fnm></au>
    <au><snm>Kim</snm><fnm>YE</fnm></au>
  </aug>
  <source>Proceedings of the 16th {ISMIR} {Conference}</source>
  <publisher>Malaga, Spain</publisher>
  <pubdate>2015</pubdate>
  <fpage>7</fpage>
</bibl>

<bibl id="B6">
  <title><p>Fashion and art cycles are driven by counter-dominance signals of
  elite competition: quantitative evidence from music styles</p></title>
  <aug>
    <au><snm>Klimek</snm><fnm>P</fnm></au>
    <au><snm>Kreuzbauer</snm><fnm>R</fnm></au>
    <au><snm>Thurner</snm><fnm>S</fnm></au>
  </aug>
  <source>Journal of The Royal Society Interface</source>
  <pubdate>2019</pubdate>
  <volume>16</volume>
  <issue>151</issue>
  <fpage>20180731</fpage>
  <url>https://royalsocietypublishing.org/doi/10.1098/rsif.2018.0731</url>
</bibl>

<bibl id="B7">
  <title><p>Leveraging the structure of musical preference in content-aware
  music recommendation</p></title>
  <aug>
    <au><snm>Magron</snm><fnm>P</fnm></au>
    <au><snm>Févotte</snm><fnm>C</fnm></au>
  </aug>
  <source>CoRR</source>
  <pubdate>2020</pubdate>
  <volume>abs/2010.10276</volume>
  <url>https://arxiv.org/abs/2010.10276</url>
  <note>arXiv: 2010.10276</note>
</bibl>

<bibl id="B8">
  <title><p>An {Evaluation} of {Audio} {Feature} {Extraction}
  {Toolboxes}</p></title>
  <aug>
    <au><snm>Moffat</snm><fnm>D</fnm></au>
    <au><snm>Ronan</snm><fnm>D</fnm></au>
    <au><snm>Reiss</snm><fnm>JD</fnm></au>
  </aug>
  <source>Proc. of the 18th {Int}. {Conference} on {Digital} {Audio} {Effects}
  ({DAFx}-15)</source>
  <publisher>Trondheim, Norway</publisher>
  <pubdate>2015</pubdate>
  <fpage>7</fpage>
</bibl>

<bibl id="B9">
  <title><p>Using perceptually defined music features in music information
  retrieval</p></title>
  <aug>
    <au><snm>Friberg</snm><fnm>A</fnm></au>
    <au><snm>Schoonderwaldt</snm><fnm>E</fnm></au>
    <au><snm>Hedblad</snm><fnm>A</fnm></au>
    <au><snm>Fabiani</snm><fnm>M</fnm></au>
    <au><snm>Elowsson</snm><fnm>A</fnm></au>
  </aug>
  <source>arXiv:1403.7923 [cs]</source>
  <pubdate>2014</pubdate>
  <url>http://arxiv.org/abs/1403.7923</url>
</bibl>

<bibl id="B10">
  <title><p>Large-{Scale} {Pattern} {Discovery} in {Music}</p></title>
  <aug>
    <au><snm>Bertin Mahieux</snm><fnm>T</fnm></au>
  </aug>
  <source>PhD thesis</source>
  <publisher>Columbia University</publisher>
  <pubdate>2013</pubdate>
  <url>https://doi.org/10.7916/D8NC67CT</url>
</bibl>

<bibl id="B11">
  <title><p>A comparison of human and automatic musical genre
  classification</p></title>
  <aug>
    <au><snm>Lippens</snm><fnm>S.</fnm></au>
    <au><snm>Martens</snm><fnm>J.P.</fnm></au>
    <au><snm>De Mulder</snm><fnm>T.</fnm></au>
  </aug>
  <source>2004 {IEEE} {International} {Conference} on {Acoustics}, {Speech},
  and {Signal} {Processing}</source>
  <publisher>Montreal, QC, Canada</publisher>
  <pubdate>2004</pubdate>
  <fpage>iv</fpage>
  <lpage>-iv</lpage>
  <note>ISSN: 1520-6149</note>
</bibl>

<bibl id="B12">
  <title><p>Musical trends and predictability of success in contemporary songs
  in and out of the top charts</p></title>
  <aug>
    <au><snm>Interiano</snm><fnm>M</fnm></au>
    <au><snm>Kazemi</snm><fnm>K</fnm></au>
    <au><snm>Wang</snm><fnm>L</fnm></au>
    <au><snm>Yang</snm><fnm>J</fnm></au>
    <au><snm>Yu</snm><fnm>Z</fnm></au>
    <au><snm>Komarova</snm><fnm>NL</fnm></au>
  </aug>
  <source>Royal Society Open Science</source>
  <pubdate>2018</pubdate>
  <volume>5</volume>
  <issue>5</issue>
  <fpage>171274</fpage>
  <url>http://royalsocietypublishing.org/doi/full/10.1098/rsos.171274</url>
  <note>Publisher: Royal Society</note>
</bibl>

<bibl id="B13">
  <title><p>Comparing {Lyrics} {Features} for {Genre} {Recognition}</p></title>
  <aug>
    <au><snm>Mayerl</snm><fnm>M</fnm></au>
    <au><snm>Votter</snm><fnm>M</fnm></au>
    <au><snm>Zangerle</snm><fnm>MME</fnm></au>
  </aug>
  <source>Proceedings of the 1st {Workshop} on {NLP} for {Music} and {Audio}
  ({NLP4MusA})</source>
  <publisher>Online: Association for Computational Linguistics</publisher>
  <pubdate>2020</pubdate>
  <fpage>73</fpage>
  <lpage>-77</lpage>
  <url>https://www.aclweb.org/anthology/2020.nlp4musa-1.15.pdf</url>
</bibl>

<bibl id="B14">
  <title><p>Combination of audio and lyrics features for genre classification
  in digital audio collections</p></title>
  <aug>
    <au><snm>Mayer</snm><fnm>R</fnm></au>
    <au><snm>Neumayer</snm><fnm>R</fnm></au>
    <au><snm>Rauber</snm><fnm>A</fnm></au>
  </aug>
  <source>Proceedings of the 16th {ACM} international conference on
  {Multimedia}</source>
  <publisher>New York, NY, USA: Association for Computing Machinery</publisher>
  <series><title><p>{MM} '08</p></title></series>
  <pubdate>2008</pubdate>
  <fpage>159</fpage>
  <lpage>-168</lpage>
  <url>https://doi.org/10.1145/1459359.1459382</url>
</bibl>

<bibl id="B15">
  <title><p>Lyric {Text} {Mining} in {Music} {Mood}
  {Classification}</p></title>
  <aug>
    <au><snm>Hu</snm><fnm>X</fnm></au>
    <au><snm>Downie</snm><fnm>JS</fnm></au>
    <au><snm>Ehmann</snm><fnm>AF</fnm></au>
  </aug>
  <source>Proceedings of the 10th {International} {Society} for {Music}
  {Information} {Retrieval} {Conference}</source>
  <publisher>Kobe International Conference Center, Kobe, Japan</publisher>
  <pubdate>2009</pubdate>
  <fpage>6</fpage>
</bibl>

<bibl id="B16">
  <title><p>Improving mood classification in music digital libraries by
  combining lyrics and audio</p></title>
  <aug>
    <au><snm>Hu</snm><fnm>X</fnm></au>
    <au><snm>Downie</snm><fnm>JS</fnm></au>
  </aug>
  <source>Proceedings of the 10th annual joint conference on {Digital}
  libraries</source>
  <publisher>New York, NY, USA: Association for Computing Machinery</publisher>
  <series><title><p>{JCDL} '10</p></title></series>
  <pubdate>2010</pubdate>
  <fpage>159</fpage>
  <lpage>-168</lpage>
  <url>https://doi.org/10.1145/1816123.1816146</url>
</bibl>

<bibl id="B17">
  <title><p>Lyric document embeddings for music tagging</p></title>
  <aug>
    <au><snm>McVicar</snm><fnm>M</fnm></au>
    <au><snm>Giorgi</snm><fnm>BD</fnm></au>
    <au><snm>Dundar</snm><fnm>B</fnm></au>
    <au><snm>Mauch</snm><fnm>M</fnm></au>
  </aug>
  <source>Proc. of the 15th {International} {Symposium} on {CMMR}</source>
  <publisher>Online</publisher>
  <pubdate>2021</pubdate>
  <fpage>10</fpage>
</bibl>

<bibl id="B18">
  <title><p>Efficient {Estimation} of {Word} {Representations} in {Vector}
  {Space}</p></title>
  <aug>
    <au><snm>Mikolov</snm><fnm>T</fnm></au>
    <au><snm>Chen</snm><fnm>K</fnm></au>
    <au><snm>Corrado</snm><fnm>G</fnm></au>
    <au><snm>Dean</snm><fnm>J</fnm></au>
  </aug>
  <source>arXiv:1301.3781 [cs]</source>
  <pubdate>2013</pubdate>
  <url>http://arxiv.org/abs/1301.3781</url>
  <note>arXiv: 1301.3781</note>
</bibl>

<bibl id="B19">
  <title><p>Distributed {Representations} of {Sentences} and
  {Documents}</p></title>
  <aug>
    <au><snm>Le</snm><fnm>QV</fnm></au>
    <au><snm>Mikolov</snm><fnm>T</fnm></au>
  </aug>
  <source>arXiv:1405.4053 [cs]</source>
  <pubdate>2014</pubdate>
  <url>http://arxiv.org/abs/1405.4053</url>
  <note>arXiv: 1405.4053</note>
</bibl>

<bibl id="B20">
  <title><p>Patent {Similarity} {Data} and {Innovation} {Metrics}</p></title>
  <aug>
    <au><snm>Whalen</snm><fnm>R</fnm></au>
    <au><snm>Lungeanu</snm><fnm>A</fnm></au>
    <au><snm>DeChurch</snm><fnm>L</fnm></au>
    <au><snm>Contractor</snm><fnm>N</fnm></au>
  </aug>
  <source>Journal of Empirical Legal Studies</source>
  <pubdate>2020</pubdate>
  <volume>17</volume>
  <issue>3</issue>
  <fpage>615</fpage>
  <lpage>-639</lpage>
  <url>https://onlinelibrary.wiley.com/doi/abs/10.1111/jels.12261</url>
  <note>eprint:
  https://onlinelibrary.wiley.com/doi/pdf/10.1111/jels.12261</note>
</bibl>

<bibl id="B21">
  <title><p>What {Makes} {Popular} {Culture} {Popular}? {Product} {Features}
  and {Optimal} {Differentiation} in {Music}</p></title>
  <aug>
    <au><snm>Askin</snm><fnm>N</fnm></au>
    <au><snm>Mauskapf</snm><fnm>M</fnm></au>
  </aug>
  <source>American Sociological Review</source>
  <pubdate>2017</pubdate>
  <volume>82</volume>
  <issue>5</issue>
  <fpage>910</fpage>
  <lpage>-944</lpage>
  <url>http://journals.sagepub.com/doi/10.1177/0003122417728662</url>
</bibl>

<bibl id="B22">
  <title><p>Cultural {Attributes} and their {Influence} on {Consumption}
  {Patterns} in {Popular} {Music}</p></title>
  <aug>
    <au><snm>Askin</snm><fnm>N</fnm></au>
    <au><snm>Mauskapf</snm><fnm>M</fnm></au>
  </aug>
  <source>(eds) {Social} {Informatics}. {SocInfo} 2014</source>
  <publisher>Cham: Springer International Publishing</publisher>
  <editor>Aiello, Luca Maria and McFarland, Daniel</editor>
  <series><title><p>Lecture {Notes} in {Computer}
  {Science}</p></title></series>
  <pubdate>2014</pubdate>
  <volume>8851</volume>
  <fpage>508</fpage>
  <lpage>-530</lpage>
  <url>https://doi.org/10.1007/978-3-319-13734-6$\_$36</url>
</bibl>

<bibl id="B23">
  <title><p>Are {Atypical} {Things} {More} {Popular}?</p></title>
  <aug>
    <au><snm>Berger</snm><fnm>J</fnm></au>
    <au><snm>Packard</snm><fnm>G</fnm></au>
  </aug>
  <source>Psychological Science</source>
  <pubdate>2018</pubdate>
  <volume>29</volume>
  <issue>7</issue>
  <fpage>1178</fpage>
  <lpage>-1184</lpage>
  <url>https://doi.org/10.1177/0956797618759465</url>
  <note>Publisher: SAGE Publications Inc</note>
</bibl>

<bibl id="B24">
  <title><p>Algorithmic {Effects} on the {Diversity} of {Consumption} on
  {Spotify}</p></title>
  <aug>
    <au><snm>Anderson</snm><fnm>A</fnm></au>
    <au><snm>Maystre</snm><fnm>L</fnm></au>
    <au><snm>Anderson</snm><fnm>I</fnm></au>
    <au><snm>Mehrotra</snm><fnm>R</fnm></au>
    <au><snm>Lalmas</snm><fnm>M</fnm></au>
  </aug>
  <source>Proceedings of {The} {Web} {Conference} 2020</source>
  <publisher>New York, NY, USA: Association for Computing Machinery</publisher>
  <pubdate>2020</pubdate>
  <fpage>2155</fpage>
  <lpage>-2165</lpage>
  <url>https://doi.org/10.1145/3366423.3380281</url>
</bibl>

<bibl id="B25">
  <title><p>Multimodal {Music} {Mood} {Classification} {Using} {Audio} and
  {Lyrics}</p></title>
  <aug>
    <au><snm>Laurier</snm><fnm>C</fnm></au>
    <au><snm>Grivolla</snm><fnm>J</fnm></au>
    <au><snm>Herrera</snm><fnm>P</fnm></au>
  </aug>
  <source>2008 {Seventh} {International} {Conference} on {Machine} {Learning}
  and {Applications}</source>
  <pubdate>2008</pubdate>
  <fpage>688</fpage>
  <lpage>-693</lpage>
</bibl>

<bibl id="B26">
  <title><p>Integration of {Text} and {Audio} {Features} for {Genre}
  {Classification} in {Music} {Information} {Retrieval}</p></title>
  <aug>
    <au><snm>Neumayer</snm><fnm>R</fnm></au>
    <au><snm>Rauber</snm><fnm>A</fnm></au>
  </aug>
  <source>Advances in {Information} {Retrieval}</source>
  <publisher>Berlin, Heidelberg: Springer Berlin Heidelberg</publisher>
  <editor>Amati, Giambattista and Carpineto, Claudio and Romano,
  Giovanni</editor>
  <series><title><p>Lecture {Notes} in {Computer}
  {Science}</p></title></series>
  <pubdate>2007</pubdate>
  <volume>4425</volume>
  <fpage>724</fpage>
  <lpage>-727</lpage>
  <url>http://link.springer.com/10.1007/978-3-540-71496-5$\_$78</url>
</bibl>

<bibl id="B27">
  <title><p>Toward automated discovery of artistic influence</p></title>
  <aug>
    <au><snm>Saleh</snm><fnm>B</fnm></au>
    <au><snm>Abe</snm><fnm>K</fnm></au>
    <au><snm>Arora</snm><fnm>RS</fnm></au>
    <au><snm>Elgammal</snm><fnm>A</fnm></au>
  </aug>
  <source>Multimedia Tools and Applications</source>
  <pubdate>2016</pubdate>
  <volume>75</volume>
  <issue>7</issue>
  <fpage>3565</fpage>
  <lpage>-3591</lpage>
  <url>http://link.springer.com/10.1007/s11042-014-2193-x</url>
</bibl>

<bibl id="B28">
  <title><p>Atypical {Combinations} and {Scientific} {Impact}</p></title>
  <aug>
    <au><snm>Uzzi</snm><fnm>B</fnm></au>
    <au><snm>Mukherjee</snm><fnm>S</fnm></au>
    <au><snm>Stringer</snm><fnm>M</fnm></au>
    <au><snm>Jones</snm><fnm>B</fnm></au>
  </aug>
  <source>Science</source>
  <pubdate>2013</pubdate>
  <volume>342</volume>
  <issue>6157</issue>
  <fpage>468</fpage>
  <lpage>-472</lpage>
  <url>https://www.science.org/doi/10.1126/science.1240474</url>
</bibl>

<bibl id="B29">
  <title><p>Analyzing information resources that support the creative
  process</p></title>
  <aug>
    <au><snm>Li</snm><fnm>Y</fnm></au>
    <au><snm>Zhang</snm><fnm>Y</fnm></au>
    <au><snm>Capra</snm><fnm>R</fnm></au>
  </aug>
  <source>{ACM} {SIGIR} {Conference} on {Human} {Information} {Interaction} and
  {Retrieval}</source>
  <publisher>New York, NY, USA: Association for Computing Machinery</publisher>
  <series><title><p>{CHIIR} '22</p></title></series>
  <pubdate>2022</pubdate>
  <fpage>180</fpage>
  <lpage>-190</lpage>
  <url>https://doi.org/10.1145/3498366.3505817</url>
</bibl>

<bibl id="B30">
  <title><p>Hot streaks in artistic, cultural, and scientific
  careers</p></title>
  <aug>
    <au><snm>Liu</snm><fnm>L</fnm></au>
    <au><snm>Wang</snm><fnm>Y</fnm></au>
    <au><snm>Sinatra</snm><fnm>R</fnm></au>
    <au><snm>Giles</snm><fnm>CL</fnm></au>
    <au><snm>Song</snm><fnm>C</fnm></au>
    <au><snm>Wang</snm><fnm>D</fnm></au>
  </aug>
  <source>Nature</source>
  <pubdate>2018</pubdate>
  <volume>559</volume>
  <issue>7714</issue>
  <fpage>396</fpage>
  <lpage>-399</lpage>
  <url>http://www.nature.com/articles/s41586-018-0315-8</url>
  <note>Number: 7714 Publisher: Nature Publishing Group</note>
</bibl>

<bibl id="B31">
  <title><p>Scientific collaboration, research funding, and novelty in
  scientific knowledge</p></title>
  <aug>
    <au><snm>Shin</snm><fnm>H</fnm></au>
    <au><snm>Kim</snm><fnm>K</fnm></au>
    <au><snm>Kogler</snm><fnm>DF</fnm></au>
  </aug>
  <source>PLOS ONE</source>
  <pubdate>2022</pubdate>
  <volume>17</volume>
  <issue>7</issue>
  <fpage>e0271678</fpage>
  <url>https://journals.plos.org/plosone/article?id=10.1371/journal.pone.0271678</url>
  <note>Publisher: Public Library of Science</note>
</bibl>

<bibl id="B32">
  <title><p>Weaving the fabric of science: {Dynamic} network models of
  science's unfolding structure</p></title>
  <aug>
    <au><snm>Shi</snm><fnm>F</fnm></au>
    <au><snm>Foster</snm><fnm>JG</fnm></au>
    <au><snm>Evans</snm><fnm>JA</fnm></au>
  </aug>
  <source>Social Networks</source>
  <pubdate>2015</pubdate>
  <volume>43</volume>
  <fpage>73</fpage>
  <lpage>-85</lpage>
  <url>http://www.sciencedirect.com/science/article/pii/S0378873315000118</url>
</bibl>

<bibl id="B33">
  <title><p>What to {Expect} {When} the {Unexpected} {Becomes} {Expected}:
  {Harmonic} {Surprise} and {Preference} {Over} {Time} in {Popular}
  {Music}</p></title>
  <aug>
    <au><snm>Miles</snm><fnm>SA</fnm></au>
    <au><snm>Rosen</snm><fnm>DS</fnm></au>
    <au><snm>Barry</snm><fnm>S</fnm></au>
    <au><snm>Grunberg</snm><fnm>D</fnm></au>
    <au><snm>Grzywacz</snm><fnm>N</fnm></au>
  </aug>
  <source>Frontiers in Human Neuroscience</source>
  <pubdate>2021</pubdate>
  <volume>15</volume>
  <url>https://www.frontiersin.org/article/10.3389/fnhum.2021.578644</url>
</bibl>

<bibl id="B34">
  <title><p>Quantitative analysis of the evolution of novelty in cinema through
  crowdsourced keywords</p></title>
  <aug>
    <au><snm>Sreenivasan</snm><fnm>S</fnm></au>
  </aug>
  <source>Scientific Reports</source>
  <pubdate>2013</pubdate>
  <volume>3</volume>
  <issue>1</issue>
  <fpage>2758</fpage>
  <url>https://www.nature.com/articles/srep02758</url>
  <note>Number: 1 Publisher: Nature Publishing Group</note>
</bibl>

<bibl id="B35">
  <title><p>Sameness {Attracts}, {Novelty} {Disturbs}, but {Outliers}
  {Flourish} in {Fanfiction} {Online}</p></title>
  <aug>
    <au><snm>Jing</snm><fnm>E</fnm></au>
    <au><snm>DeDeo</snm><fnm>S</fnm></au>
    <au><snm>Ahn</snm><fnm>YY</fnm></au>
  </aug>
  <source>arXiv:1904.07741 [cs]</source>
  <pubdate>2019</pubdate>
  <url>http://arxiv.org/abs/1904.07741</url>
  <note>arXiv: 1904.07741</note>
</bibl>

<bibl id="B36">
  <title><p>Novelty and influence of creative works, and quantifying patterns
  of advances based on probabilistic references networks</p></title>
  <aug>
    <au><snm>Park</snm><fnm>D</fnm></au>
    <au><snm>Nam</snm><fnm>J</fnm></au>
    <au><snm>Park</snm><fnm>J</fnm></au>
  </aug>
  <source>EPJ Data Science</source>
  <pubdate>2020</pubdate>
  <volume>9</volume>
  <issue>1</issue>
  <fpage>1</fpage>
  <lpage>-15</lpage>
  <url>https://epjdatascience.springeropen.com/articles/10.1140/epjds/s13688-019-0214-8</url>
  <note>Number: 1 Publisher: SpringerOpen</note>
</bibl>

<bibl id="B37">
  <title><p>Pandemics are catalysts of scientific novelty: {Evidence} from
  {COVID}-19</p></title>
  <aug>
    <au><snm>Liu</snm><fnm>M</fnm></au>
    <au><snm>Bu</snm><fnm>Y</fnm></au>
    <au><snm>Chen</snm><fnm>C</fnm></au>
    <au><snm>Xu</snm><fnm>J</fnm></au>
    <au><snm>Li</snm><fnm>D</fnm></au>
    <au><snm>Leng</snm><fnm>Y</fnm></au>
    <au><snm>Freeman</snm><fnm>RB</fnm></au>
    <au><snm>Meyer</snm><fnm>ET</fnm></au>
    <au><snm>Yoon</snm><fnm>W</fnm></au>
    <au><snm>Sung</snm><fnm>M</fnm></au>
    <au><snm>Jeong</snm><fnm>M</fnm></au>
    <au><snm>Lee</snm><fnm>J</fnm></au>
    <au><snm>Kang</snm><fnm>J</fnm></au>
    <au><snm>Min</snm><fnm>C</fnm></au>
    <au><snm>Song</snm><fnm>M</fnm></au>
    <au><snm>Zhai</snm><fnm>Y</fnm></au>
    <au><snm>Ding</snm><fnm>Y</fnm></au>
  </aug>
  <source>Journal of the Association for Information Science and
  Technology</source>
  <pubdate>2022</pubdate>
  <volume>73</volume>
  <issue>8</issue>
  <fpage>1065</fpage>
  <lpage>-1078</lpage>
  <url>https://onlinelibrary.wiley.com/doi/abs/10.1002/asi.24612</url>
  <note>\_eprint:
  https://onlinelibrary.wiley.com/doi/pdf/10.1002/asi.24612</note>
</bibl>

<bibl id="B38">
  <title><p>Exploring {Acoustic} {Similartiy} for {Novel} {Music}
  {Recommendation}</p></title>
  <aug>
    <au><snm>Cheng</snm><fnm>D</fnm></au>
    <au><snm>Joachims</snm><fnm>T</fnm></au>
    <au><snm>Turnbull</snm><fnm>D</fnm></au>
  </aug>
  <pubdate>2020</pubdate>
  <fpage>7</fpage>
</bibl>

<bibl id="B39">
  <title><p>Hit song prediction: {Leveraging} low- and high-level audio
  features</p></title>
  <aug>
    <au><snm>Zangerle</snm><fnm>E</fnm></au>
    <au><snm>Huber</snm><fnm>R</fnm></au>
    <au><snm>Vötter</snm><fnm>M</fnm></au>
    <au><snm>Yang</snm><fnm>YH</fnm></au>
  </aug>
  <source>Proceedings of the 20th International Society for Music Information
  Retrieval Conference, ISMIR 2019</source>
  <pubdate>2019</pubdate>
  <fpage>319</fpage>
  <lpage>-326</lpage>
  <url>https://doi.org/10.5281/zenodo.3258042</url>
</bibl>

<bibl id="B40">
  <title><p>Taste {Over} {Time}: {The} {Temporal} {Dynamics} of {User}
  {Preferences}</p></title>
  <aug>
    <au><snm>Moore</snm><fnm>JL</fnm></au>
    <au><snm>Chen</snm><fnm>S</fnm></au>
    <au><snm>Joachims</snm><fnm>T</fnm></au>
    <au><snm>Turnbull</snm><fnm>D</fnm></au>
  </aug>
  <source>Proceedings of the 14th International Society for Music Information
  Retrieval Conference, ISMIR 2013</source>
  <pubdate>2013</pubdate>
  <fpage>6</fpage>
</bibl>

<bibl id="B41">
  <title><p>Novelty, complexity, and hedonic value</p></title>
  <aug>
    <au><snm>Berlyne</snm><fnm>D. E.</fnm></au>
  </aug>
  <source>Perception \& Psychophysics</source>
  <pubdate>1970</pubdate>
  <volume>8</volume>
  <issue>5</issue>
  <fpage>279</fpage>
  <lpage>-286</lpage>
  <url>https://doi.org/10.3758/BF03212593</url>
</bibl>

<bibl id="B42">
  <title><p>Back to the inverted-{U} for music preference: {A} review of the
  literature</p></title>
  <aug>
    <au><snm>Chmiel</snm><fnm>A</fnm></au>
    <au><snm>Schubert</snm><fnm>E</fnm></au>
  </aug>
  <source>Psychology of Music</source>
  <pubdate>2017</pubdate>
  <volume>45</volume>
  <issue>6</issue>
  <fpage>886</fpage>
  <lpage>-909</lpage>
  <url>http://journals.sagepub.com/doi/10.1177/0305735617697507</url>
</bibl>

<bibl id="B43">
  <title><p>Breakthrough recognition: {Bias} against novelty and competition
  for attention</p></title>
  <aug>
    <au><snm>Chai</snm><fnm>S</fnm></au>
    <au><snm>Menon</snm><fnm>A</fnm></au>
  </aug>
  <source>Research Policy</source>
  <pubdate>2019</pubdate>
  <volume>48</volume>
  <issue>3</issue>
  <fpage>733</fpage>
  <lpage>-747</lpage>
  <url>https://www.sciencedirect.com/science/article/abs/pii/S0048733318302774</url>
</bibl>

<bibl id="B44">
  <title><p>Bias against novelty in science: {A} cautionary tale for users of
  bibliometric indicators</p></title>
  <aug>
    <au><snm>Wang</snm><fnm>J</fnm></au>
    <au><snm>Veugelers</snm><fnm>R</fnm></au>
    <au><snm>Stephan</snm><fnm>P</fnm></au>
  </aug>
  <source>Research Policy</source>
  <pubdate>2017</pubdate>
  <volume>46</volume>
  <issue>8</issue>
  <fpage>1416</fpage>
  <lpage>-1436</lpage>
  <url>https://ideas.repec.org//a/eee/respol/v46y2017i8p1416-1436.html</url>
  <note>Publisher: Elsevier</note>
</bibl>

<bibl id="B45">
  <title><p>Gensim: topic modelling for humans</p></title>
  <aug>
    <au><cnm>{Radim Řehůřek}</cnm></au>
    <au><snm>Sojka</snm><fnm>P</fnm></au>
  </aug>
  <source>Proceedings of the {LREC} 2010 {Workshop} on {New} {Challenges} for
  {NLP} {Frameworks}</source>
  <publisher>Valletta, Malta: ELRA</publisher>
  <pubdate>2010</pubdate>
  <fpage>45</fpage>
  <lpage>-50</lpage>
  <url>https://radimrehurek.com/gensim/models/doc2vec.html</url>
</bibl>

<bibl id="B46">
  <title><p>Scikit-learn: {Machine} {Learning} in {Python}</p></title>
  <aug>
    <au><snm>Pedregosa</snm><fnm>F</fnm></au>
    <au><snm>Varoquaux</snm><fnm>G</fnm></au>
    <au><snm>Gramfort</snm><fnm>A</fnm></au>
    <au><snm>Michel</snm><fnm>V</fnm></au>
    <au><snm>Thirion</snm><fnm>B</fnm></au>
    <au><snm>Grisel</snm><fnm>O</fnm></au>
    <au><snm>Blondel</snm><fnm>M</fnm></au>
    <au><snm>Prettenhofer</snm><fnm>P</fnm></au>
    <au><snm>Weiss</snm><fnm>R</fnm></au>
    <au><snm>Dubourg</snm><fnm>V</fnm></au>
    <au><snm>Vanderplas</snm><fnm>J</fnm></au>
    <au><snm>Passos</snm><fnm>A</fnm></au>
    <au><snm>Cournapeau</snm><fnm>D</fnm></au>
    <au><snm>Brucher</snm><fnm>M</fnm></au>
    <au><snm>Perrot</snm><fnm>M</fnm></au>
    <au><snm>Duchesnay</snm><fnm>E</fnm></au>
  </aug>
  <source>Journal of Machine Learning Research</source>
  <pubdate>2011</pubdate>
  <volume>12</volume>
  <issue>85</issue>
  <fpage>2825</fpage>
  <lpage>-2830</lpage>
  <url>http://jmlr.org/papers/v12/pedregosa11a.html</url>
</bibl>

<bibl id="B47">
  <title><p>Singing in the {Brain}: {Independence} of {Lyrics} and
  {Tunes}</p></title>
  <aug>
    <au><snm>Besson</snm><fnm>M.</fnm></au>
    <au><snm>Faïta</snm><fnm>F.</fnm></au>
    <au><snm>Peretz</snm><fnm>I.</fnm></au>
    <au><snm>Bonnel</snm><fnm>A. M.</fnm></au>
    <au><snm>Requin</snm><fnm>J.</fnm></au>
  </aug>
  <source>Psychological Science</source>
  <pubdate>1998</pubdate>
  <volume>9</volume>
  <issue>6</issue>
  <fpage>494</fpage>
  <lpage>-498</lpage>
  <url>https://doi.org/10.1111/1467-9280.00091</url>
  <note>Publisher: SAGE Publications Inc</note>
</bibl>

<bibl id="B48">
  <title><p>Early selectivity for vocal and musical sounds:
  electrophysiological evidence from an adaptation paradigm</p></title>
  <aug>
    <au><snm>Rigoulot</snm><fnm>S</fnm></au>
    <au><snm>Armony</snm><fnm>JL</fnm></au>
  </aug>
  <source>European Journal of Neuroscience</source>
  <pubdate>2016</pubdate>
  <volume>44</volume>
  <issue>10</issue>
  <fpage>2786</fpage>
  <lpage>-2794</lpage>
  <url>https://onlinelibrary.wiley.com/doi/abs/10.1111/ejn.13391</url>
  <note>\_eprint:
  https://onlinelibrary.wiley.com/doi/pdf/10.1111/ejn.13391</note>
</bibl>

<bibl id="B49">
  <title><p>Exposure effects on music preference and recognition</p></title>
  <aug>
    <au><snm>Peretz</snm><fnm>I</fnm></au>
    <au><snm>Gaudreau</snm><fnm>D</fnm></au>
    <au><snm>Bonnel</snm><fnm>AM</fnm></au>
  </aug>
  <source>Memory \& Cognition</source>
  <pubdate>1998</pubdate>
  <volume>26</volume>
  <issue>5</issue>
  <fpage>884</fpage>
  <lpage>-902</lpage>
  <url>https://doi.org/10.3758/BF03201171</url>
</bibl>

<bibl id="B50">
  <title><p>Novelty and collective attention</p></title>
  <aug>
    <au><snm>Wu</snm><fnm>F</fnm></au>
    <au><snm>Huberman</snm><fnm>BA</fnm></au>
  </aug>
  <source>Proceedings of the National Academy of Sciences</source>
  <pubdate>2007</pubdate>
  <volume>104</volume>
  <issue>45</issue>
  <fpage>17599</fpage>
  <lpage>-17601</lpage>
  <url>https://www.pnas.org/doi/10.1073/pnas.0704916104</url>
  <note>Publisher: Proceedings of the National Academy of Sciences</note>
</bibl>

<bibl id="B51">
  <title><p>Experimental {Study} of {Inequality} and {Unpredictability} in an
  {Artificial} {Cultural} {Market}</p></title>
  <aug>
    <au><snm>Salganik</snm><fnm>MJ</fnm></au>
    <au><snm>Dodds</snm><fnm>PS</fnm></au>
    <au><snm>Watts</snm><fnm>DJ</fnm></au>
  </aug>
  <source>Science</source>
  <pubdate>2006</pubdate>
  <volume>311</volume>
  <issue>5762</issue>
  <fpage>854</fpage>
  <lpage>-856</lpage>
  <url>https://science.sciencemag.org/content/311/5762/854</url>
  <note>Publisher: American Association for the Advancement of Science Section:
  Report</note>
</bibl>

<bibl id="B52">
  <title><p>Things {Change}: {Comparing} {Results} {Using} {Historical} {Data}
  and {User} {Testing} for {Evaluating} a {Recommendation} {Task}</p></title>
  <aug>
    <au><snm>Jung</snm><fnm>SG</fnm></au>
    <au><snm>Salminen</snm><fnm>J</fnm></au>
    <au><snm>Chowdhury</snm><fnm>SA</fnm></au>
    <au><snm>Ramirez Robillos</snm><fnm>D</fnm></au>
    <au><snm>Jansen</snm><fnm>BJ</fnm></au>
  </aug>
  <source>Extended {Abstracts} of the 2020 {CHI} {Conference} on {Human}
  {Factors} in {Computing} {Systems}</source>
  <publisher>New York, NY, USA: Association for Computing Machinery</publisher>
  <series><title><p>{CHI} {EA} '20</p></title></series>
  <pubdate>2020</pubdate>
  <fpage>1</fpage>
  <lpage>-7</lpage>
  <url>https://doi.org/10.1145/3334480.3382945</url>
</bibl>

<bibl id="B53">
  <title><p>Enhancing {Collaborative} {Filtering} {Music} {Recommendation} by
  {Balancing} {Exploration} and {Exploitation}</p></title>
  <aug>
    <au><snm>Xing</snm><fnm>Z</fnm></au>
    <au><snm>Wang</snm><fnm>X</fnm></au>
    <au><snm>Wang</snm><fnm>Y</fnm></au>
  </aug>
  <source>Proceedings of the 15th {International} {Society} for {Music}
  {Information} {Retrieval} {Conference}</source>
  <publisher>Taipei, Taiwan</publisher>
  <pubdate>2014</pubdate>
</bibl>

<bibl id="B54">
  <title><p>The {Wisdom} of the {Few}? “{Supertaggers}” in {Collaborative}
  {Tagging} {Systems}</p></title>
  <aug>
    <au><snm>Lorince</snm><fnm>J</fnm></au>
    <au><snm>Zorowitz</snm><fnm>S</fnm></au>
    <au><snm>Murdock</snm><fnm>J</fnm></au>
    <au><snm>Todd</snm><fnm>PM</fnm></au>
  </aug>
  <source>The Journal of Web Science</source>
  <pubdate>2015</pubdate>
  <volume>1</volume>
  <url>https://webscience-journal.net/webscience/article/view/12</url>
</bibl>

</refgrp>
} 









\section*{Appendix}
\begin{table}[ht]
\centering
\caption{MIR extracted audio features provided by The Echo Nest API}
\label{table:Echo_Nest}
 \begin{tabular}{p{0.2\linewidth} | p{0.7\linewidth}} 
\textbf{Audio Feature} & \textbf{Value Description}
 \\ \hline
duration\textunderscore ms &	The duration of the track in milliseconds.
 \\ \hline
key 	&	The estimated overall key of the track. Integers map to pitches using standard Pitch Class notation . E.g. 0 = C, 1 = C$\sharp$/D$\flat$, 2 = D, and so on. If no key was detected, the value is -1.
 \\  \hline
mode 	&	Mode indicates the modality (major or minor) of a track. Major is represented by 1 and minor is 0.
 \\ \hline
time\textunderscore signature & 	An estimated overall time signature of a track. 
 \\    \hline
acousticness 	& 	A confidence measure from 0.0 to 1.0 of whether the track is acoustic. 1.0 represents high confidence the track is acoustic.
 \\    \hline
danceability 	&	Danceability describes how suitable a track is for dancing based on a combination of musical elements.
 \\    \hline
energy 	& 	Energy is a measure from 0.0 to 1.0 and represents a perceptual measure of intensity and activity. 
\\    \hline
instrumentalness 	& 	Predicts whether a track contains no vocals. The closer the instrumentalness value is to 1.0, the greater likelihood the track contains no vocal content. 
\\    \hline
liveness  &	 Higher liveness values represent an increased probability that the track was performed live. 
\\    \hline
loudness 	&	The overall loudness of a track in decibels (dB). 
\\    \hline
speechiness 	& 	Speechiness detects the presence of spoken words in a track. The more exclusively speech-like the recording (e.g. talk show, audio book, poetry), the closer to 1.0 the attribute value. 
\\    \hline
valence 	&	A measure from 0.0 to 1.0 describing the musical positiveness conveyed by a track. 
\\    \hline
tempo 	&	The overall estimated tempo of a track in beats per minute (BPM). 
\\    \hline

\end{tabular}

\end{table}

\subsection*{Influence Probability Calculations}

For the sake of reducing noise, we did not use the overall average of the relative novelty change for the aggregated influence probability value. Instead, two dummy variables were created, one corresponding with lyric relative novelty change, and one with MIR relative novelty change. For each song, if its change in lyric relative novelty was negative, the corresponding dummy variable was assigned a 1, and if positive, a 0. The same was done for MIR relative novelty change. When grouping the songs by novelty range and total weeks on chart, we then took the average of the appropriate dummy variable to calculate the probability that a song with those parameters would have a decrease in relative novelty, indicating a higher likelihood of being influential. Because the distribution of data meant that the size of each sample varied for the different numbers of total weeks on chart, the influence probability over the range of total weeks was calculated using a weighted rolling window average, with a window size of 8. Additionally, because there are a relatively few songs that spend more than 40 weeks on the chart, almost all of which fall into the optimal novelty bin for both modalities, we limited our analysis to the 0-35 week range.




\end{backmatter}
\end{document}